%% file: root_arXiv.tex
\documentclass[10pt,journal,,twoside]{IEEEtran}
\input{preamble.tex}
\def\BibTeX{{\rm B\kern-.05em{\sc i\kern-.025em b}\kern-.08em
    T\kern-.1667em\lower.7ex\hbox{E}\kern-.125emX}}
\title{How many autonomous vehicles are required to stabilize traffic flow?}

\author{MirSaleh Bahavarnia$^\dagger$, \IEEEmembership{Member, IEEE} and Ahmad F. Taha$^\dagger$, \IEEEmembership{Member, IEEE}
	\thanks{$^\dagger$The authors are with the Department of Civil and Environmental Engineering, Vanderbilt University, 2201 West End Avenue, TN 37235, USA. Ahmad F. Taha is also affiliated with the Department of Electrical and Computer Engineering.} 
\thanks{Emails: \{mirsaleh.bahavarnia,ahmad.taha\}@vanderbilt.edu.}
}

\begin{document}

\maketitle

\begin{abstract}
The collective behavior of \textit{human-driven vehicles} (HVs) produces the well-known stop-and-go waves potentially leading to higher fuel consumption and emissions. This paper investigates the stabilization of traffic flow via a minimum number of \textit{autonomous vehicles} (AVs) subject to constraints on the control parameters aiming to reduce the number of vehicles on the road while achieving lower fuel consumption and emissions. The unconstrained scenario has been well-studied in recent studies. The main motivation to investigate the constrained scenario is that, in realistic engineering applications, lower and upper bounds exist on the control parameters. For the constrained scenario, we optimally find the minimum number of required AVs (via computing the optimal lower bound on the AV penetration rate) to stabilize traffic flow for a given number of HVs. As an immediate consequence, we conclude that for a given number of AVs, the number of HVs in the stabilized traffic flow may not be arbitrarily large in the constrained scenario unlike the unconstrained scenario studied in the literature. We systematically propose a procedure to compute the optimal lower bound on the AV penetration rate using nonlinear optimization techniques. Finally, we validate the theoretical results via numerical simulations. Numerical simulations suggest that enlarging the constraint intervals makes a smaller optimal lower bound on the AV penetration rate attainable. However, it leads to a slower transient response due to a dominant pole closer to the origin.
\end{abstract}

\begin{IEEEkeywords}
Autonomous vehicles, Constrained control, Stability of linear systems, Traffic control, Transportation networks.
\end{IEEEkeywords}

\section{Introduction and Paper Contributions} \label{sec:Intro}

\IEEEPARstart{T}{he} collective behavior of \textit{human-driven vehicles} (HVs) produces the well-known stop-and-go waves potentially leading to undesirable higher vehicle fuel consumption and emissions. Thus, the stabilization of traffic flow via \textit{autonomous vehicles} (AVs) has attained great attention in traffic flow control \cite{cui2017stabilizing,stern2018dissipation,wu2018stabilizing,zheng2020smoothing,wang2021controllability,wang2021optimal,wang2023optimal,wang2023general,ameli2024designing,zhou2020stabilizing,bahavarnia2024on} as it can significantly smooth the stop-and-go waves and improve the efficiency of vehicle fuel consumption and emissions. In \cite{wilson2011car}, developing a general framework for car-following models, various linear stability concepts (e.g., string stability \cite{swaroop1996string}) along with the corresponding linear stability analyses are detailed. For a linearized car-following model around a uniform flow equilibrium state, string stability is equivalent to the system with no increasing (i.e., unstable) eigenmodes, that is, the linearized dynamics is said to be string stable if infinitesimal perturbations do not amplify and the system remains close to the equilibrium \cite{wilson2011car,cui2017stabilizing}.

In \cite{cui2017stabilizing,wu2018stabilizing}, the authors have shown that via a single AV, in the absence of noise (ideal circumstance), traffic flow can be stabilized. In an experimental study conducted by \cite{stern2018dissipation}, it is experimentally verified that a single AV can control the flow of $20$ HVs around it, with significant reductions in velocity standard deviation, excessive braking, and fuel consumption. Considering the \textit{optimal-velocity} (OV) model \cite{bando1995dynamical}, the authors in \cite{zheng2020smoothing,wang2021controllability}, have proved that the mixed vehicular platoon consisting of a single AV and multiple HVs is not completely controllable, but is stabilizable and synthesized $\mathcal{H}_2$ optimal state feedback controller to actively mitigate undesirable traffic perturbations. Built upon general nonlinear car-following dynamics, the authors in \cite{wang2021optimal} have formulated an optimal control problem (Bolza problem) aiming to minimize vehicle speed perturbation. Following \cite{wang2021optimal}, taking advantage of a min-max approach, they have derived an optimal feedback control law for AVs in the presence of cyber-attacks in \cite{wang2023optimal}. In \cite{wang2023general}, based on the general functional form of car-following dynamics, the authors have proposed effective additive AV controllers with provable speed profile tracking convergence along with safety and string stability enabling sufficient conditions. In \cite{chou2022lord}, a thorough comparative analysis has been conducted among ten AV algorithms in the literature in terms of diverse performance metrics like time to stabilize, maximum headway, vehicle miles traveled, and fuel economy. A comprehensive literature review of AV control can be found in \cite{di2021survey}. In a recent thorough experimental study \cite{ameli2024designing}, a live traffic control experiment involving $100$ vehicles near Nashville, Tennessee was conducted to implement various controllers to smooth stop-and-go traffic waves. In that study, AVs were simulated in multiple scenarios to evaluate their effect on traffic congestion. In \cite{wu2021flow}, leveraging deep reinforcement learning (RL) methods, the authors have presented a modular learning framework to improve the quality of traffic congestion alleviation via RL-based controllers compared to the model-based alternatives and to potentially equip the real-time advisory (RTA) systems accordingly \cite{hasan2024lessons}.

On one hand, in the absence of noise (ideal circumstance), traffic flow can be stabilized via a single AV by employing a high-gain controller with a sufficiently high gain \cite{cui2017stabilizing} while in a more realistic scenario, controller gains are constrained by the lower and upper bounds affecting the speed of the transient response. Although the authors in \cite{wu2018stabilizing} have considered a bounded search space of the physically realizable AV controller gains solely for the numerical simulations, the corresponding lower and upper bounds of the bounded search space and the rational driving constraints (RDC) \cite{wilson2011car} are not systematically incorporated into the theoretical setup of the optimization problem. On the other hand, by systematically incorporating the lower and upper bounds on the controller gains, the authors in \cite{bahavarnia2024on} have proposed a constrained version of an unconstrained CAV platoon $\mathcal{H}_{\infty}$ optimal controller synthesis \cite{zhou2020stabilizing} with an ultimate application to the mixed vehicular platoons consisting of both HVs and AVs. 

\noindent \textbf{Research Question.} Motivated by the high-gain controller limitation in \cite{cui2017stabilizing} and taking into account a more realistic gain-constrained scenario similar to the one considered by \cite{bahavarnia2024on}, one can pose the following two-part question for a mixed vehicular platoon consisting of both HVs and AVs:

\textit{Q1: (i) Can we systematically stabilize traffic flow via AVs subject to the lower and upper bounds on the control parameters? (ii) If the answer is yes, what is the minimum number of required AVs to that end?}

\noindent \textbf{Paper Contributions.} Throughout this paper, considering the second-order car-following model utilized by \cite{cui2017stabilizing}, we aim to answer \textit{Q1} thoroughly.
The main contributions of the paper can be listed as follows:

\begin{itemize}
    \item Considering a circular road with a single lane, no ramps, and uniform conditions, we consider a constrained version (incorporating the lower and upper bounds on the control parameters) of an unconstrained problem on stabilizing traffic flow via autonomous vehicles \cite{cui2017stabilizing} aiming to answer \textit{Q1}.
    \item The theoretical contributions are threefold: \textit{(i)} derivation of necessary and sufficient conditions for the string stability criterion, \textit{(ii)} parameterization of the rational driving constraints (RDC) \cite{wilson2011car} and the box constraints (encoding the lower and upper bounds on the control parameters), and \textit{(iii)} derivation of the optimal lower bound on the AV penetration rate.
    \item We present a procedure to compute the optimal lower bound on the AV penetration rate using nonlinear optimization. As an immediate consequence, for a given number of AVs, the number of HVs in the stabilized traffic flow may not be arbitrarily large in the constrained scenario unlike the unconstrained scenario studied in \cite{cui2017stabilizing}. Finally, we validate the theoretical results via numerical simulations. Numerical simulations suggest that enlarging the constraint intervals makes a smaller optimal lower bound on the AV penetration rate attainable. However, it leads to a slower transient response due to a dominant pole closer to the origin.
    \item We also introduce an $\mathcal{H}_{\infty}$-based measure to quantify the string stability quality and employ a greedy algorithm to obtain a sub-optimal distribution of the AVs to reduce the string stability conservatism.
\end{itemize}

\noindent \textbf{Paper Organization.} The remainder of the paper is structured as follows: Section \ref{VDPS} details the vehicle dynamics (both human and autonomous vehicles) and states the problem to be studied. Section \ref{SAV} elaborates on \textit{(i)} deriving necessary and sufficient conditions for the string stability criterion, \textit{(ii)} constructing parameterization of the RDC and box constraints, and \textit{(iii)} deriving the optimal lower bound on the AV penetration rate. Section \ref{Prcre} presents a procedure to compute the optimal lower bound on the AV penetration rate followed by the numerical simulations to assess the validity of the theoretical results. Finally, Section \ref{Con} concludes the paper with a few concluding remarks including some future directions. Paper notation is presented in Appendix \ref{App1}.

\section{Vehicles Dynamics and Problem Statement} \label{VDPS}

We consider a circular road with a single lane, no ramps, and uniform road conditions. The main reasons for this choice are as follows \cite{cui2017stabilizing}: \textit{(i)} it does not require the consideration of boundary conditions, \textit{(ii)} it corresponds to an infinite straight road setup (i.e., $n \to \infty$ and scaling $L \propto n$) with $\frac{1}{\gamma}$-periodic traffic dynamics ($n$: Number of vehicles, $L$: Road length, and $\gamma := \frac{m}{n}$: AV penetration rate, $m$: Number of AVs), \textit{(iii)} it has a record of experimental instabilities \cite{sugiyama2008traffic} enabling the calibration of model parameters, and \textit{(iv)} it facilitates the theoretical derivations due to the periodicity. Tab. \ref{tab:my_label} in Appendix \ref{App2} summarizes traffic flow dynamics quantities. Let us assume the following ordering of the vehicles: vehicle $j+1$ precedes (leads) vehicle $j$ for $j \in \mathbb{N}_n$ (for $j = n$, vehicle $n+1$ is defined as vehicle $1$). In this paper, similar to \cite{cui2017stabilizing}, we limit our attention to the case of near-equilibrium flow (i.e., local stabilization). Then, for such near-equilibrium flow, collision avoidance is automatically resolved by ensuring the local string stability subject to small perturbations \cite{cui2017stabilizing} and systematically incorporating the rational driving constraints (RDC) \cite{wilson2011car}. Vehicles can be categorized into two types: \textit{(i)} human-driven vehicles (HVs), and \textit{(ii)} autonomous vehicles (AVs). Then, we accordingly have $\mathbb{N}_n = \mathcal{I}_{\mathrm{HV}} \cup \mathcal{I}_{\mathrm{AV}}$ with $|\mathcal{I}_{\mathrm{HV}}| = n-m$ and $|\mathcal{I}_{\mathrm{AV}}| = m$.

\subsection{HVs dynamics}

For each HV, we consider the following second-order car-following dynamics:
\begin{align} \label{HVE}
    \ddot{x}_j(t) &= f(h_j(t),\dot{h}_j(t),v_j(t)),~j \in \mathcal{I}_{\mathrm{HV}}.
\end{align}Among many examples, one important example of a car-following dynamics describable by \eqref{HVE}, is the \textit{optimal-velocity-follow-the-leader} (OV-FTL) model \cite{bando1994structure,cui2017stabilizing}.

Considering dynamics \eqref{HVE} under small perturbations from the equilibrium flow, we obtain the following linearized dynamics:
{\small \begin{subequations} \label{LHVE}
\begin{align}
    & \ddot{y}_j(t) = \alpha_1 (y_{j+1}(t)-y_j(t)) - \alpha_2 u_j(t) + \alpha_3 u_{j+1}(t), \label{LHa}\\
    &\alpha_1 = \frac{\partial f}{\partial h_j} \bigg{|}_{\mathrm{eq}},\alpha_2 = \frac{\partial f}{\partial \dot{h}_j}\bigg{|}_{\mathrm{eq}}-\frac{\partial f}{\partial v_j}\bigg{|}_{\mathrm{eq}},\alpha_3 = \frac{\partial f}{\partial \dot{h}_j}\bigg{|}_{\mathrm{eq}}, \label{LHb}\\
    & j \in \mathcal{I}_{\mathrm{HV}}.
\end{align}   
\end{subequations}}For linearized dynamics \eqref{LHVE}, $\forall j \in \mathcal{I}_{\mathrm{HV}}$, the following standard assumptions hold \cite{cui2017stabilizing}: the acceleration of vehicle $j$ is reduced when \textit{(i)} the spacing $h_j(t)$ decreases, \textit{(ii)} the relative velocity $\dot{h}_j(t)$ decreases, or \textit{(iii)} the vehicle's velocity $v_j(t)$ increases. Such risk aversion criteria imply the following rational driving constraints (RDC) \cite{wilson2011car}: $\alpha_1 > 0$, $\alpha_2 > \alpha_3$, and $\alpha_3 > 0$. To determine the poles associated with linearized dynamics \eqref{LHVE}, one can take the Laplace transformation from \eqref{LHVE} leading to the following transfer function:
{\small \begin{subequations} \label{TFHV}
\begin{align}
    & T_j(s) := \frac{Y_j(s)}{Y_{j+1}(s)} = F(s;\alpha) = \frac{\alpha_3 s + \alpha_1}{s^2 + \alpha_2 s + \alpha_1},~j \in \mathcal{I}_{\mathrm{HV}},
\end{align}    
\end{subequations}}where $\alpha := \begin{bmatrix}
    \alpha_1 & \alpha_2 & \alpha_3
\end{bmatrix}^\top$ denotes the system parameters vector. Remarkably, the Hurwitz stability of transfer function $F(s;\alpha)$ is equivalent to the simultaneous satisfaction of $\alpha_1 > 0$ and $\alpha_2 > 0$.

\subsection{AVs dynamics}

Similarly, for each AV, we consider the following second-order car-following dynamics:
\begin{align} \label{AVE}
    \ddot{x}_j(t) &= g(h_j(t),\dot{h}_j(t),v_j(t)),~j \in \mathcal{I}_{\mathrm{AV}}.
\end{align}Considering dynamics \eqref{AVE} under small perturbations from the equilibrium flow, we obtain the following linearized dynamics:
{\small \begin{subequations} \label{LAVE}
\begin{align}
    & \ddot{y}_j(t) = \beta_1 (y_{j+1}(t)-y_j(t)) - \beta_2 u_j(t) + \beta_3 u_{j+1}(t), \label{LAa}\\
    &\beta_1 = \frac{\partial g}{\partial h_j} \bigg{|}_{\mathrm{eq}},\beta_2 = \frac{\partial g}{\partial \dot{h}_j}\bigg{|}_{\mathrm{eq}}-\frac{\partial g}{\partial v_j}\bigg{|}_{\mathrm{eq}},\beta_3 = \frac{\partial g}{\partial \dot{h}_j}\bigg{|}_{\mathrm{eq}}, \label{LAb}\\
    & j \in \mathcal{I}_{\mathrm{AV}}.
\end{align}   
\end{subequations}}Likewise, for linearized dynamics \eqref{LAVE}, $\forall j \in \mathcal{I}_{\mathrm{AV}}$, the following standard assumptions hold \cite{cui2017stabilizing}: the acceleration of vehicle $j$ is reduced when \textit{(i)} the spacing $h_j(t)$ decreases, \textit{(ii)} the relative velocity $\dot{h}_j(t)$ decreases, or \textit{(iii)} the vehicle's velocity $v_j(t)$ increases. Such risk aversion criteria imply the following RDC \cite{wilson2011car}:
\begin{align} \label{SignEqs}
    & \beta_1 > 0,~\beta_2 - \beta_3 > 0,~\beta_3 > 0.
\end{align}Similarly, to determine the poles associated with linearized dynamics \eqref{LAVE}, one can take the Laplace transformation from \eqref{LAVE} leading to the following transfer function:
{\small \begin{subequations} \label{TFAV}
\begin{align}
    & T_j(s) := \frac{Y_j(s)}{Y_{j+1}(s)} = G(s;\beta) = \frac{\beta_3 s + \beta_1}{s^2 + \beta_2 s + \beta_1},~ j \in \mathcal{I}_{\mathrm{AV}},
\end{align}    
\end{subequations}}where $\beta := \begin{bmatrix}
    \beta_1 & \beta_2 & \beta_3
\end{bmatrix}^\top$ denotes the control parameters vector. Remarkably, the Hurwitz stability of transfer function $G(s;\beta)$ is equivalent to the simultaneous satisfaction of $\beta_1 > 0$ and $\beta_2 > 0$.

\subsection{Problem statement}

One can obtain the linearized dynamics associated with the mixed vehicular platoon consisting of both HVs and AVs by simply augmenting the HVs linearized dynamics \eqref{LHVE} and AVs linearized dynamics \eqref{LAVE} as follows:
\begin{align} \label{LMVP}
\mathrm{Linearized~dynamics} &: \begin{cases}
\eqref{LHa},\eqref{LHb}, & j \in \mathcal{I}_{\mathrm{HV}}\\
\eqref{LAa},\eqref{LAb}, & j \in \mathcal{I}_{\mathrm{AV}}
    \end{cases}.
\end{align}
A mixed vehicular platoon with linearized dynamics \eqref{LMVP} is said to be \textit{string stable} if infinitesimal perturbations do not amplify and the system remains close to the equilibrium \cite{wilson2011car,cui2017stabilizing}. The formal mathematical definition of string stability can be expressed as follows:
\begin{mydef}[String Stability] \label{def1}
    A mixed vehicular platoon with linearized dynamics \eqref{LMVP} is said to be \textit{string stable} if all of its eigenmodes lie on the left half plane $\mathbb{C}^{-} := \{s \in \mathbb{C}: \Re(s) \le 0\}$.
\end{mydef}

Let us assume the following lower and upper bounds on the control parameters vector $\beta$:
\begin{subequations} \label{LUB}
\begin{align} 
    &\beta_1^l \le \beta_1 \le \beta_1^u,~\beta_2^l \le \beta_2 \le \beta_2^u,~\beta_3^l \le \beta_3 \le \beta_3^u.
\end{align}    
\end{subequations}Note that $0 < \beta_i^l$ holds for all $i \in \{1,2,3\}$. Similarly, we use the following notations: $\beta^{l} := \begin{bmatrix}
    \beta_1^{l} & \beta_2^{l} & \beta_3^{l}
\end{bmatrix}^\top$ and $\beta^{u} := \begin{bmatrix}
    \beta_1^{u} & \beta_2^{u} & \beta_3^{u}
\end{bmatrix}^\top$ in the sequel where needed. Throughout this paper, we mathematically investigate the following problem:

\begin{mypbm} \label{Prob1}
    Given a mixed vehicular platoon with linearized dynamics \eqref{LMVP}, the RDC \eqref{SignEqs}, and the lower and upper bounds \eqref{LUB} on the control parameters vector $\beta$, find the optimal $\beta^{\ast}$ for which traffic flow can be stabilized with an optimally minimum AV penetration rate.
\end{mypbm}

\section{String Stability with AVs} \label{SAV}

This section is comprised of the following three main parts: \textit{(i)} derivation of necessary and sufficient conditions for the string stability criterion, \textit{(ii)} parameterization of the RDC and the box constraints (encoding the lower and upper bounds on the control parameters), and \textit{(iii)} derivation of the optimal lower bound on the AV penetration rate.

\subsection{Necessary and sufficient conditions for the string stability criterion}

Considering \eqref{TFHV} and \eqref{TFAV}, and according to the periodicity of the circular road, we have
\begin{align} \label{PCR}
    \prod_{j \in \mathbb{N}_n} T_j(s) &= F(s;\alpha)^{n-m} G(s;\beta)^m = 1.
\end{align}
Note that the eigenmodes of linearized dynamics \eqref{LMVP} are the $2n$ roots of \eqref{PCR}. The $2n$ roots of \eqref{PCR} lie in a curve  $\mathcal{C} := \{s \in \mathbb{C}: |F(s;\alpha)|^{1-\gamma} |G(s;\beta)|^{\gamma} = 1\}$ where $\gamma := \frac{m}{n}$ denotes the AV penetration rate. A sufficient string stability condition can be formulated as $\mathcal{C} \subset \mathbb{C}^{-}$. Let us define the following fractional-order transfer function:
\begin{align} \label{Hgam}
    & H_{\gamma}(s) := F(s;\alpha)^{1-\gamma}G(s;\beta)^\gamma.
\end{align}To ensure the string stability of linearized dynamics \eqref{LMVP}, it suffices to consider $\mathcal{C} \subset \mathbb{C}^{-}$ and equivalently impose $|H_{\gamma}(\imath \omega)| \le 1$ for all $\omega \in \mathbb{R}$ (i.e., $\|H_{\gamma}(s)\|_{\infty} \le 1$) which is equivalent to the following string stability criterion \cite{cui2017stabilizing}:

{\small \begin{subequations} \label{logfor}
\begin{align}
    & (1-\gamma)D_{\alpha}(\omega) + \gamma D_{\beta}(\omega) \le 0,~\forall \omega \in \mathbb{R}, \label{logA}\\
    & \hspace{-0.025in}D_{\alpha}(\omega) := \ln (|F(\imath \omega;\alpha)|) = \frac{1}{2} \ln \bigg (\frac{\alpha_3^2 \omega^2 + \alpha_1^2}{\alpha_2^2 \omega^2 + (\omega^2-\alpha_1)^2} \bigg), \label{logB}\\
    & D_{\beta}(\omega) := \ln (|G(\imath \omega;\beta)|) = \frac{1}{2} \ln \bigg (\frac{\beta_3^2 \omega^2 + \beta_1^2}{\beta_2^2 \omega^2 + (\omega^2-\beta_1)^2} \bigg). \label{logC}
\end{align}   
\end{subequations}}As stated by \cite{cui2017stabilizing}, we have
\begin{subequations} \label{FGDel}
\begin{align}
    & |F(\imath \omega;\alpha)| \le 1,~\forall \omega \in \mathbb{R} \iff \Delta_\alpha \ge 0,\label{FDel}\\
    & \Delta_\alpha := -2\alpha_1 + \alpha_2^2 - \alpha_3^2, \label{DelF}\\
    & |G(\imath \omega;\beta)| \le 1,~\forall \omega \in \mathbb{R} \iff \Delta_\beta \ge 0,\label{GDel}\\
    & \Delta_\beta := -2\beta_1 + \beta_2^2 - \beta_3^2.\label{DelG}
\end{align}   
\end{subequations}Note that equivalences $|F(\imath \omega;\alpha)| \le 1 \iff D_\alpha(\omega) \le 0$ and $|G(\imath \omega;\beta)| \le 1 \iff D_\beta(\omega) \le 0$ simply hold as we respectively have $D_{\alpha}(\omega) := \ln (|F(\imath \omega;\alpha)|)$ and $D_{\beta}(\omega) := \ln (|G(\imath \omega;\beta)|)$.

Without loss of generality, we can only consider the case of $\omega \ge 0$ as $\omega^2 = (-\omega)^2$ holds. One can verify that in the case of $\Delta_\alpha < 0$, we have
\begin{align} \label{SignD}
    \mathrm{Sign}~\mathrm{of}~D_\alpha(\omega) &: \begin{cases}
        D_\alpha(\omega) > 0, & \omega \in (0,\sqrt{-\Delta_\alpha})\\
        D_\alpha(\omega) = 0, & \omega \in \{0,\sqrt{-\Delta_\alpha}\}\\
        D_\alpha(\omega) < 0, & \omega \in (\sqrt{-\Delta_\alpha},\infty)
    \end{cases}.
\end{align}
Now, we are ready to state the main result of this paper.

\begin{myprs} \label{Propo1}
    Consider the worst-case scenario for which the HVs dynamics violate $|F(\imath \omega;\alpha)| \le 1$ for some $\omega \in \mathbb{R}$. The string stability criterion \eqref{logfor} holds for $\beta$ if and only if
    \begin{subequations} \label{IffEq}
        \begin{align}
        & \Delta_{\beta} \ge 0,\label{DelB} \\
        & \gamma \ge \frac{1}{J^\ast(\beta) + 1},~J^\ast(\beta) := \inf \{J(\omega;\beta): \omega \in \mathbb{I}_{\alpha}\}, \label{gam1}
    \end{align}  
    \end{subequations}
    hold for $\beta$ where $\mathbb{I}_{\alpha}$ and $J(\omega;\beta)$ denote $(0,\sqrt{-\Delta_\alpha})$ and $-\frac{D_{\beta}(\omega)}{D_{\alpha}(\omega)}$, respectively. Moreover, $J^{\ast}(\beta)$ in \eqref{gam1} can be computed as follows:
    \begin{subequations} \label{Jbet}
        \begin{align}
        & J^{\ast}(\beta) = \min \bigg \{\frac{\alpha_1^2}{-\Delta_{\alpha}} \frac{\Delta_{\beta}}{\beta_1^2}, J(\omega;\beta) |_{\omega \in \mathcal{J}}\bigg \}, \label{Jb} \\
        & \mathcal{J} := \bigg \{\omega \in \mathbb{I}_{\alpha}: \frac{dJ(\omega;\beta)}{d\omega} = 0, \frac{d^2J(\omega;\beta)}{d\omega^2} \ge 0 \bigg \}. \label{CalJ}
    \end{align}    
    \end{subequations}
\end{myprs}
\begin{IEEEproof}
    See Appendix \ref{App3}.
\end{IEEEproof} 
It is noteworthy that merging \eqref{gam1} and $\gamma < 1$ implies that the necessary condition $J^{\ast}(\beta) > 0$ must hold for any $\beta$ satisfying the stability condition \eqref{DelB}. We emphasize that $J^{\ast}(\beta)$ in \eqref{gam1} is only defined for $\beta$s satisfying the stability condition \eqref{DelB}.

\subsection{Parameterization of the RDC and box constraints}

Let us define the following notations: $\mathcal{B}_1 := \{\beta \in \mathbb{R}^3: \eqref{SignEqs}~\mathrm{holds}~\mathrm{for}~\beta\}$, $\mathcal{B}_2 := \{\beta \in \mathbb{R}^3: \eqref{DelB}~\mathrm{holds}~\mathrm{for}~\beta\}$, and $\mathcal{B}_3 := \{\beta \in \mathbb{R}^3: \eqref{LUB}~\mathrm{holds}~\mathrm{for}~\beta\}$. Proposition \ref{Propo1} expresses the necessary and sufficient conditions on $\beta$ and $\gamma$ to ensure the string stability criterion \eqref{logfor} holds. Specifically, it states $\beta \in \mathcal{B}_2$ must hold. However, we need to additionally impose $\beta \in \mathcal{B}_1 \cap \mathcal{B}_3$ according to the setup in Problem \ref{Prob1} as the RDC \eqref{SignEqs} and the lower and upper bounds \eqref{LUB} on the control parameters $\begin{bmatrix}
    \beta_1 & \beta_2 & \beta_3
\end{bmatrix}$ must be satisfied. The set of the control parameters $\begin{bmatrix}
    \beta_1 & \beta_2 & \beta_3
\end{bmatrix}$ satisfying the RDC \eqref{SignEqs} and the string stability condition \eqref{DelB}, i.e., $\mathcal{B}_1 \cap \mathcal{B}_2$, can be parameterized via the parameters $\begin{bmatrix}
    p & q & r
\end{bmatrix}$ as
{\small \begin{align} \label{pqr}
    & \beta_3(p) = p,~\beta_2(p,q) = p + q,~\beta_1(p,q,r) = pq + \frac{q^2}{2} - r,
\end{align}}where $p > 0$, $q > 0$, and $r \ge 0$ hold.

In the following proposition, we systematically incorporate the box constraints \eqref{LUB} into the parameterization \eqref{pqr}.

\begin{myprs} \label{Propo2}
    The parameters $\begin{bmatrix}
    p & q & r
\end{bmatrix}$ in \eqref{pqr} satisfying box constraints \eqref{LUB} can be parameterized via the parameters $\begin{bmatrix}
    \psi_1 & \psi_2 & \psi_3
\end{bmatrix}$ as
    {\small \begin{subequations} \label{ParPar}
        \begin{align}
            & p = \texttt{p}(\psi_1) = (1-\psi_1)p^l + \psi_1 p^u,\\
            & q = \texttt{q}(\psi_1,\psi_2) = (1-\psi_2)q_{\psi_1}^l + \psi_2 q_{\psi_1}^u,\\
            & r = \texttt{r}(\psi_1,\psi_2,\psi_3) = (1-\psi_3)r_{\psi_1,\psi_2}^l + \psi_3 r_{\psi_1,\psi_2}^u,\\
            & \textrm{with} \notag\\
            & p^l = \max\{\epsilon,\beta_3^l\},\\
            & p^u = \min \Big \{\beta_3^u,\beta_2^u-\epsilon,\sqrt{{\beta_2^u}^2-2\beta_1^l} \Big\},\\
            & q_{\psi_1}^l = \max \Big \{\epsilon,\beta_2^l-\texttt{p}(\psi_1),\sqrt{\texttt{p}(\psi_1)^2+2\beta_1^l}-\texttt{p}(\psi_1)\Big \},\\ 
            & q_{\psi_1}^u = \beta_2^u - \texttt{p}(\psi_1),\\
            & r_{\psi_1,\psi_2}^l \hspace{-0.025in} = \max \bigg \{0,\texttt{p}(\psi_1) \texttt{q}(\psi_1,\psi_2) + \frac{\texttt{q}(\psi_1,\psi_2)^2}{2}-\beta_1^u \bigg\}, \\
            & r_{\psi_1,\psi_2}^u = \texttt{p}(\psi_1) \texttt{q}(\psi_1,\psi_2) + \frac{\texttt{q}(\psi_1,\psi_2)^2}{2}-\beta_1^l,
        \end{align}
    \end{subequations}}where $\psi_i \in [0,1]$ holds for all $i \in \{1,2,3\}$ and $\epsilon > 0$ is an infinitesimal value. Moreover, $\beta_3^u \ge \epsilon$ and $\beta_2^u \ge \max \Big \{2\epsilon,\beta_3^l + \epsilon, \sqrt{{\beta_3^l}^2+2\beta_1^l} \Big\}$ must hold as necessary conditions for $\beta^l$ and $\beta^u$. 
\end{myprs}
\begin{IEEEproof}
    See Appendix \ref{App4}.
\end{IEEEproof} 
In \eqref{ParPar}, we can opt the form of $\psi_i$s via an arbitrary sigmoid function, e.g., the logistic function
\begin{align} \label{sigfun}
    \phi(\tau) &= 1/(1 + e^{-\zeta \tau}), 
\end{align}
where $\zeta > 0$ represents the logistic growth rate. Such a choice is reasonable as we need a one-to-one mapping between $(-\infty,\infty)$ and $(0,1)$ for parameterization. Some other sigmoid function choices for the $\phi(\tau)$ can be built upon the hyperbolic tangent function $\mathrm{tanh}{({\texttt{x}})} = (e^{\texttt{x}}-e^{-{\texttt{x}}})/(e^{\texttt{x}}+e^{-{\texttt{x}}})$, the arctangent function $\mathrm{arctan}{({\texttt{x}})}$, and the error function $\mathrm{erf}({\texttt{x}}) := (2/\sqrt{\pi}) \int_0^{\texttt{x}} e^{-{\texttt{t}}^2} d{\texttt{t}}$ to name a few. Combining \eqref{pqr} and \eqref{ParPar} along with sigmoid function \eqref{sigfun}, we state the following corollary.
\begin{mycor} \label{cor1}
    The control parameters $\begin{bmatrix}
    \beta_1 & \beta_2 & \beta_3
\end{bmatrix}$ satisfying the RDC \eqref{SignEqs}, the string stability condition \eqref{DelB}, and the box constraints \eqref{LUB}, i.e., any member of the set $\mathcal{B}_1 \cap \mathcal{B}_2 \cap \mathcal{B}_3$, can be parameterized via the parameters $\begin{bmatrix}
    \theta_1 & \theta_2 & \theta_3
\end{bmatrix}$ as
\begin{subequations} \label{Partheta}
    \begin{align}
        & \beta_3(\theta_1) = \texttt{p}(\phi(\theta_1)),\\
        & \beta_2(\theta_1,\theta_2) = \texttt{p}(\phi(\theta_1)) + \texttt{q}(\phi(\theta_1),\phi(\theta_2)),\\
        & \beta_1(\theta_1,\theta_2,\theta_3) = \texttt{p}(\phi(\theta_1)) \texttt{q}(\phi(\theta_1),\phi(\theta_2)) \notag\\ 
        & +\frac{\texttt{q}(\phi(\theta_1),\phi(\theta_2))^2}{2} - \texttt{r}(\phi(\theta_1),\phi(\theta_2),\phi(\theta_3)),
    \end{align}
\end{subequations}where $\theta_i \in \mathbb{R}$ holds for all $i \in \{1,2,3\}$, $\phi()$ denote the logistic function in \eqref{sigfun}, and $\texttt{p}()$, $\texttt{q}()$, and $\texttt{r}()$ represent the same functions expressed in \eqref{ParPar}. 
\end{mycor}

We use the following notations: $\theta := \begin{bmatrix}
        \theta_1 & \theta_2 & \theta_3
    \end{bmatrix}^\top$ and $\beta(\theta) := \begin{bmatrix}
        \beta_1(\theta_1,\theta_2,\theta_3) & \beta_2(\theta_1,\theta_2) & \beta_3(\theta_1)
    \end{bmatrix}^\top$ in the sequel where needed.

\subsection{Optimal lower bound on the AV penetration rate}

Built upon the parameterized stabilizing control parameters vector $\beta(\theta)$ characterized by \eqref{Partheta} in Corollary \ref{cor1}, we state the following corollary.
\begin{mycor} \label{cor2}
    Consider the worst-case scenario for which the HVs dynamics violate $|F(\imath\omega;\alpha)| \le 1$ for some $\omega \in \mathbb{R}$. Solving the following optimization problem:
        \begin{align}
            & \underset{\theta \in \mathbb{R}^3}{\max}~J^{\ast}(\beta(\theta)),\label{Equ2}
        \end{align}
for $\theta^{\ast}$ and denoting the optimal parameterized stabilizing control parameters vector by $\beta(\theta^{\ast})$, the AV penetration rate is optimally lower bounded by
        \begin{align}
            & \gamma \ge 1/(J^{\ast \ast}+1),\label{Equ1}
        \end{align}
    where $J^{\ast \ast} < \infty$ denotes the optimal value associated with the optimization problem \eqref{Equ2}.
\end{mycor}

According to Corollary \ref{cor2}, we derive the optimal lower bound on the AV penetration rate to stabilize traffic flow. It is noteworthy that the optimal lower bound $\frac{1}{J^{\ast \ast}+1}$ in \eqref{Equ1} implicitly depends on the human-driven vehicle dynamics $\alpha$ and the lower and upper bounds on the control parameters $\beta^l$ and $\beta^u$ affecting the AV dynamics. Moreover, denoting the number of AVs and the number of HVs by $N_{\mathrm{AV}}$ and $N_{\mathrm{HV}}$, respectively, \eqref{Equ1} is simply equivalent to any of the following equivalent inequalities:
\begin{subequations} \label{LBUB}
    \begin{align}
    & N_{\mathrm{AV}} \ge \lceil{ {N_{\mathrm{HV}}}/{J^{\ast \ast}}} \rceil, \label{LB}\\
    & N_{\mathrm{HV}} \le \lfloor J^{\ast \ast} N_{\mathrm{AV}} \rfloor. \label{UB}
    \end{align}
\end{subequations}
Given $N_{\mathrm{HV}}$, \eqref{LB} presents the optimal minimum number of the required AVs to stabilize traffic flow. Equivalently, given $N_{\mathrm{AV}}$, \eqref{UB} presents the optimal maximum number of HVs for which traffic flow can be stabilized. As an insightful observation for the special case of $N_{\mathrm{AV}} = 1$, we observe that \eqref{UB} implies that $N_{\mathrm{HV}} \le \lfloor J^{\ast \ast} \rfloor$ should hold, that is, for a given number of AVs, the number of HVs in the stabilized traffic flow may not be arbitrarily large in the constrained scenario unlike the unconstrained scenario studied in \cite{cui2017stabilizing}. In other terms, if $N_{\mathrm{HV}} \ge \lfloor J^{\ast \ast} \rfloor + 1$ holds, then traffic flow may not be stabilized with a single AV in the constrained scenario unlike the unconstrained scenario studied in \cite{cui2017stabilizing}. 

As a summary, Corollary \ref{cor1} will be utilized as a cornerstone to characterize the set of the control parameters $\begin{bmatrix}
    \beta_1 & \beta_2 & \beta_3
\end{bmatrix}$ satisfying the RDC \eqref{SignEqs}, the string stability condition \eqref{DelB}, and the box constraints \eqref{LUB}, i.e., $\mathcal{B}_1 \cap \mathcal{B}_2 \cap \mathcal{B}_3$ via the parameters $\begin{bmatrix}
    \theta_1 & \theta_2 & \theta_3
\end{bmatrix}$. Such parameterization facilitates the computation of the optimal value $J^{\ast \ast}$ in optimization problem \eqref{Equ2} stated by Corollary \ref{cor2}. The proposed method minimizes the AV penetration rate leading to a reduced number of vehicles on the road. To achieve more traffic control objectives (e.g., demand serving efficiency) in addition, one can mathematically model the desired objective function or incorporate the corresponding constraints into the current state-space model and then solve the newly formulated optimization problem to obtain the new optimal controller gains accordingly.

\section{Numerical Simulations} \label{Prcre}
This section presents Procedure \ref{proc:Lbf} to compute the optimal lower bound on the AV penetration rate. Then, we assess the validity of the theoretical results by conducting numerical simulations in MATLAB R2024a. In Procedure \ref{proc:Lbf}, we utilize the MATLAB built-in function $\texttt{fminsearch}()$ (developed based on Nelder– Mead simplex method \cite{lagarias1998convergence}) as our nonlinear optimization solver to solve \eqref{Equ2} for $\theta^{\ast}$ along with $\theta_0$ as an initial $\theta$. Appendix \ref{App5} presents an alternative approach to compute the optimal lower bound on the AV penetration rate. Appendices \ref{App6} and \ref{App7} provide more insights on string stability conservatism and extra observations via additional numerical simulations, respectively.

\setlength{\floatsep}{5pt}{
\begin{algorithm}[!ht]
\caption{\textit{Optimal lower bound on the AV penetration rate finder}}\label{proc:Lbf}
\DontPrintSemicolon
\textbf{Input:} $\alpha$, $\beta^{l}$, $\beta^{u}$.

\hspace{0.1in} Compute $J^{\ast}(\beta)$ via \eqref{Jbet}.

\hspace{0.1in} Construct $J^{\ast}(\beta(\theta))$ via the parameterization \eqref{Partheta}.

\hspace{0.1in} Initialize $\theta$ with an initial $\theta$, namely $\theta_0$.

\hspace{0.1in} Solve \eqref{Equ2} for $\theta^{\ast}$ along with $\theta_0$ as an initial $\theta$.

\hspace{0.1in} Compute $J^{\ast \ast}$ via $J^{\ast \ast} = J^{\ast}(\beta(\theta^{\ast}))$.

\textbf{Output:} ${1}/(J^{\ast \ast} + 1)$.
\end{algorithm}
}

To test the theoretical results, built upon the utilized numerical setup associated with the OV-modeled HVs dynamics in \cite{zheng2020smoothing}, we consider the following numerical setup: $\alpha = \begin{bmatrix} 0.3 \pi & 1.5 & 0.9 \end{bmatrix}^\top$ with $\Delta_{\alpha} = -0.4450 < 0$, i.e., the worst-case scenario for which the HVs dynamics violate $|F(\imath\omega;\alpha)| \le 1$ for some $\omega \in \mathbb{R}$. Given the lower and upper bounds on the control parameters as  $\beta^l = \begin{bmatrix}
        0.01 & 0.01 & 0.01
    \end{bmatrix}^\top$ and $\beta^u = \begin{bmatrix}
        2 & 2 & 2
    \end{bmatrix}^\top$, and running Procedure \ref{proc:Lbf}, we get $\beta(\theta^{\ast}) = \begin{bmatrix}
        0.01 & 2 & 0.01
\end{bmatrix}^\top$ for which $J^{\ast \ast} = 184.9594$ and $\frac{1}{J^{\ast \ast}+1} = 0.0054$. Given $N_{\mathrm{AV}} = 1$, \eqref{UB} implies that $N_{\mathrm{HV}} \le \lfloor 184.9594 \times 1 \rfloor = 184$ should hold, that is, for a single AV, the number of HVs in the stabilized traffic flow may not be larger than $184$ in the constrained scenario unlike the unconstrained scenario studied in \cite{cui2017stabilizing}. Given $N_{\mathrm{HV}} = 400$, \eqref{LB} implies that $N_{\mathrm{AV}} \ge \lceil \frac{400}{184.9594} \rceil = 3$ should hold, that is, traffic flow may not be stabilized with a single AV in the constrained scenario unlike the unconstrained scenario studied in \cite{cui2017stabilizing}. 

\begin{figure}[H]
    \centering
    \includegraphics[scale = 0.21]{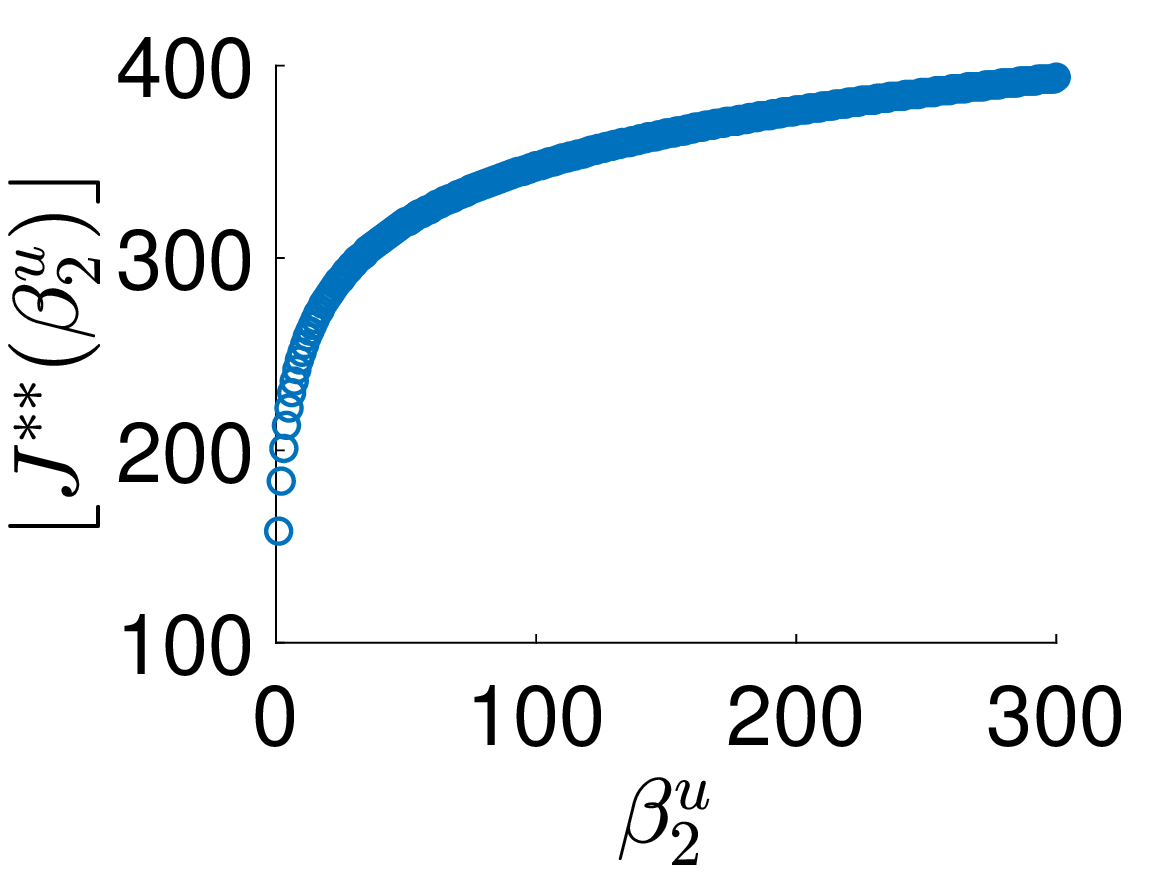}
    \includegraphics[scale = 0.21]{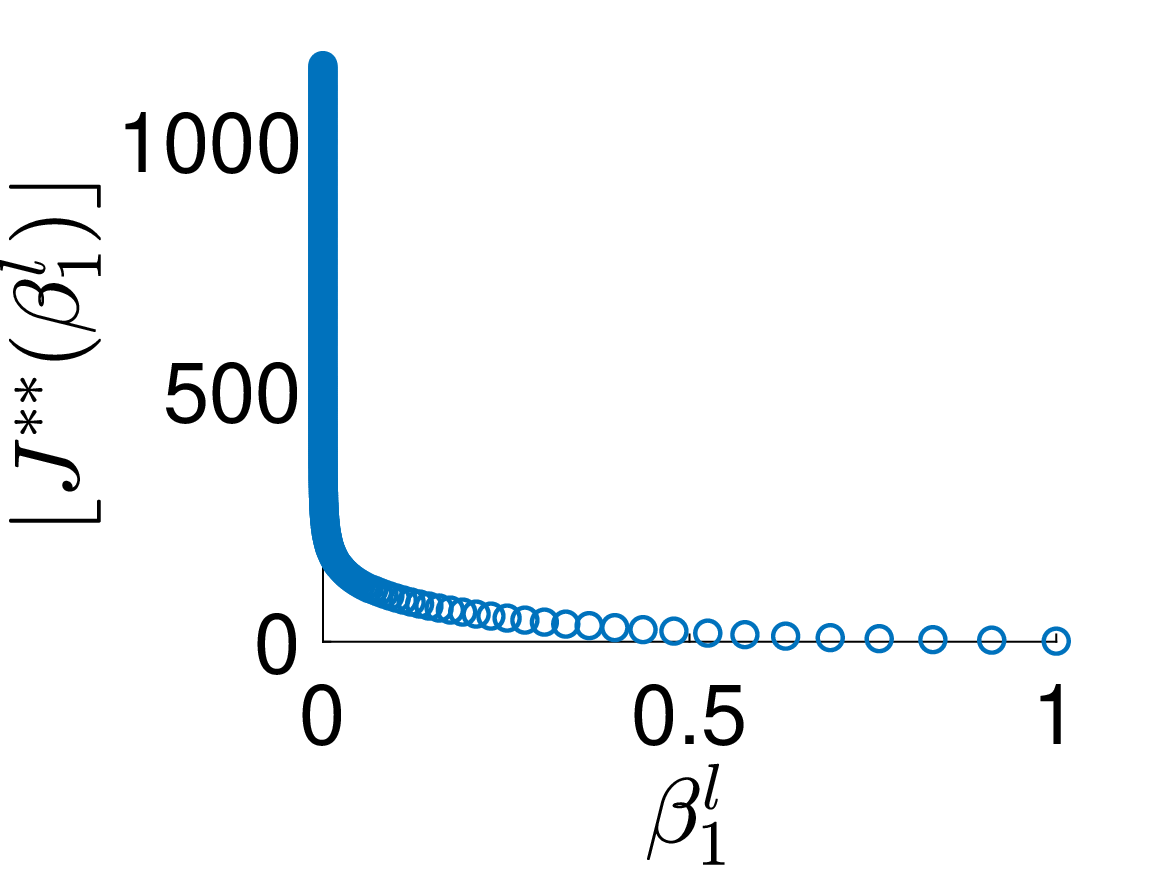}
    
    \caption{First scenario: dependency of $\lfloor J^{\ast \ast}(\beta_2^u) \rfloor$ on $\beta_2^u$ (Left). Second scenario: dependency of $\lfloor J^{\ast \ast}(\beta_1^l) \rfloor$ on $\beta_1^l$ (Right).}
    \label{fig:1}
\end{figure}

To investigate the effects of the lower and upper bounds on the control parameters \eqref{LUB} on the optimal maximum number of HVs for which traffic flow can be stabilized \eqref{UB} with a single AV, we consider the following two scenarios: \textit{(i)} $\beta^l = \begin{bmatrix}
    0.01 & 0.01 & 0.01
\end{bmatrix}^\top$, $\beta^u = \begin{bmatrix}
    i & i & i
\end{bmatrix}^\top$, $i \in \{1,\dots,300\}$, and \textit{(ii)} $\beta^l = \begin{bmatrix}
    10^{\frac{i-301}{25}} & 10^{\frac{i-301}{25}} & 10^{\frac{i-301}{25}}
\end{bmatrix}^\top$, $i \in \{1,\dots,301\}$, $\beta^u = \begin{bmatrix}
    2 & 2 & 2
\end{bmatrix}^\top$. For the first scenario, running Procedure \ref{proc:Lbf} for $i \in \{1,\dots,300\}$, we obtain Fig \ref{fig:1} (on the left) visualizing the dependency of $\lfloor J^{\ast \ast}(\beta_2^u) \rfloor$ on $\beta_2^u$. As Fig \ref{fig:1} (on the left) depicts, the larger the upper bound $\beta_2^u$, the larger number of HVs in the stabilized traffic flow can be maintained. For the second scenario, running Procedure \ref{proc:Lbf} for $i \in \{1,\dots,301\}$, we obtain Fig \ref{fig:1} (on the right) visualizing the dependency of $\lfloor J^{\ast \ast}(\beta_1^l) \rfloor$ on $\beta_1^l$. As Fig \ref{fig:1} (on the right) depicts, the smaller the lower bound $\beta_1^l$, the larger number of HVs in the stabilized traffic flow can be maintained. Although maintaining the larger number of HVs in the stabilized traffic flow is desired (for the larger values of $\beta_2^u$ in the first scenario depicted by Fig. \ref{fig:1} (on the left) and the smaller values of $\beta_1^l$ in the second scenario depicted by Fig. \ref{fig:1} (on the right)), it leads to a slower transient response due to a dominant pole $\frac{-\beta_2 \pm \sqrt{\beta_2^2-4\beta_1}}{2}$ closer to the origin. Particularly, in the second scenario, note that the smaller values of $\beta_1^l$ correspond to less risk aversion according to the RDC \eqref{SignEqs} encoding the risk aversion criteria.

\section{Concluding Remarks} \label{Con}

In this paper, we answer \textit{Q1}, posed in Section \ref{sec:Intro}, as follows: (i) The answer is yes. We can systematically stabilize traffic flow via AVs subject to the lower and upper bounds on the control parameters using nonlinear optimization techniques (as utilized in Procedure \ref{proc:Lbf}). (ii) We optimally find the minimum number of required AVs (via computing the optimal lower bound on the AV penetration rate) to stabilize traffic flow for a given number of HVs. Such optimal lower bound on the AV penetration rate implicitly depends on the HV dynamics and the lower and upper bounds on the control parameters affecting the AV dynamics. As an immediate consequence, we observe that in the case of a constrained scenario, unlike the unconstrained scenario \cite{cui2017stabilizing}, an arbitrarily large number of HVs may not be stabilized with a given number of AVs (e.g., a single AV considered in \cite{cui2017stabilizing}). In other terms, a given number of AVs (e.g., a single AV) may become insufficient to stabilize traffic flow for a sufficiently large number of HVs. The current paper is a primary research work. We highlight that a huge gap exists between the primary theoretical results and the practical situation in reality due to the idealized conditions and assumptions made through the theoretical derivations.

\textit{Future Directions}: More practical simulations with nonlinear models and large perturbations would give us a more realistic understanding of the proposed method. Then, it can be considered a pertinent practical future research direction. Another potential future direction could be generalizing the theoretical results to the case of multi-lane mixed vehicular platoons via hybrid system stability analysis techniques similar to the ones employed by \cite{li2024hybrid}.

\section*{Acknowledgments}

This work was supported by the National Science Foundation under Grants 2152450 and 2152928. The authors would like to thank Prof. Dan Work for his valuable suggestions on the numerical simulations.

\bibliographystyle{IEEEtran}
\bibliography{References}

\appendices

\section{Paper Notation} \label{App1}

We denote the Laplace domain variable by $s$. We represent the real and complex numbers sets by $\mathbb{R}$ and $\mathbb{C}$, respectively. To show the real part and absolute value of a complex number $z$, we use $\Re(z)$ and $|z|$, respectively. We use $\imath$ to denote the imaginary unit $\sqrt{-1}$. We use $\cup$ and $\cap$ to show the set union and intersection, respectively. For a set $S$, the symbols $|S|$, $\inf S$, and $\min S$, denote the cardinality, infimum, and minimum of set $S$, respectively. We show the $d$-dimensional vector of all zeros and all ones by $\mathbf{0}_d$ and $\mathbf{1}_d$, respectively. We use $\lceil . \rceil$ and $\lfloor . \rfloor$ to represent the ceiling function and the floor function, respectively. For a scalar transfer function $T(s)$, we denote its $\mathcal{H}_{\infty}$ norm by $\|T(s)\|_{\infty}$ which is defined as $\|T(s)\|_{\infty} := \underset{\omega > 0}{\sup}~|T(\imath \omega)|$. We represent the remainder after division (modulo operation) by $\mathrm{mod}(a,b)$ (when we divide $a$ by $b$).

\section{Traffic Flow Dynamics Quantities} \label{App2}

Tab. \ref{tab:my_label} summarizes traffic flow dynamics quantities.

\begin{table}[!ht]
    \centering
    \caption{Summary of traffic flow dynamics quantities}
    \begin{tabular}{|c|p{2.3in}|}
    \hline
        $\mathrm{Notation}$ & $\mathrm{Definition}$ \\
        \hline
        $L$ & $\mathrm{Road~length}$ \\
        \hline
        $n$ & $\mathrm{Number~of~vehicles}$ \\
        \hline
        $m$ & $\mathrm{Number~of~AVs}$\\
        \hline
        $n-m$ & $\mathrm{Number~of~HVs}$\\
        \hline
        $\gamma := \frac{m}{n}$ & $\mathrm{AV~penetration~rate}$\\
        \hline
        $\mathbb{N}_n$ & $\mathrm{Index~set~associated~with~vehicles}$: \newline $\{1,\dots,n\}$\\
        \hline
        $\mathcal{I}_{\mathrm{AV}}$ & $\mathrm{Index~set~associated~with~AVs}$\\
        \hline
        $\mathcal{I}_{\mathrm{HV}}$ & $\mathrm{Index~set~associated~with~HVs}$\\
        \hline
        $t$ & $\mathrm{Time}$\\
        \hline
        $x_j(t)$ & $\mathrm{Position~along~the~road~(defined~modulo}~L\mathrm{)}$ \newline $\mathrm{of~the}$~$j$-$\mathrm{th}$~$\mathrm{vehicle~at~time}~t$\\
        \hline
        $v_j(t) := \dot{x}_j(t)$ & $\mathrm{Velocity~of~the}$~$j$-$\mathrm{th}$~$\mathrm{vehicle~at~time}~t$\\
        \hline
        $a_j(t) := \ddot{x}_j(t)$ & $\mathrm{Acceleration~of~the}$~$j$-$\mathrm{th}$~$\mathrm{vehicle~at~time}~t$\\
        \hline
        $h_j(t)$ & $\mathrm{Spacing~of~the}$~$j$-$\mathrm{th}$~$\mathrm{vehicle~at~time}~t$: \newline $x_{j+1}(t)-x_j(t)$\\
        \hline
        $\dot{h}_j(t)$ & $\mathrm{Relative~velocity~of~the}$~$j$-$\mathrm{th}$~$\mathrm{vehicle~at}$ \newline $\mathrm{time}~t$: $\dot{x}_{j+1}(t)-\dot{x}_j(t)$\\
        \hline
        $x_{\mathrm{eq}}(t) \in \mathbb{R}^n$ & $\mathrm{Equilibrium~position}$\\
        \hline
        $v_{\mathrm{eq}}(t) \in \mathbb{R}^n$ & $\mathrm{Equilibrium~velocity}$\\
        \hline
        $a_{\mathrm{eq}} = \mathbf{0}_n$ & $\mathrm{Equilibrium~acceleration}$\\
        \hline
        $h_{\mathrm{eq}} = \frac{L}{n} \mathbf{1}_n$ & $\mathrm{Equilibrium~spacing}$\\
        \hline
        $y_j(t)$ & $\mathrm{Infinitesimal~position~difference~deviation~of}$ \newline $\mathrm{the}$~$j$-$\mathrm{th}$~$\mathrm{vehicle~at~time}~t$: $x_j(t)-x_{\mathrm{eq}}(t)$\\
        \hline
        $u_j(t)$ & $\mathrm{Infinitesimal~velocity~difference~deviation~of}$ \newline $\mathrm{the}$~$j$-$\mathrm{th}$~$\mathrm{vehicle~at~time}~t$: $v_j(t)-v_{\mathrm{eq}}(t)$\\
        \hline
    \end{tabular}
    
    \label{tab:my_label}
\end{table}

\section{Proof of Proposition \ref{Propo1}} \label{App3}

We consider the worst-case scenario for which the HVs dynamics violate $|F(\imath \omega;\alpha)| \le 1$ for some $\omega \in \mathbb{R}$, i.e., $|F(\imath \omega;\alpha)| > 1$ or equivalently $D_{\alpha}(\omega) > 0$ holds for some $\omega \in \mathbb{R}$. Thus, $\Delta_{\alpha} < 0$ holds according to \eqref{FDel} and \eqref{DelF}. Then, to ensure the string stability criterion \eqref{logfor} holds for $\beta$, we must have $D_{\beta}(\omega) \le 0$ for all $\omega \in \mathbb{R}$ or equivalently $\Delta_\beta \ge 0$ according to \eqref{GDel} and \eqref{DelG}. Since $\Delta_{\alpha} < 0$ holds, we have $D_{\alpha}(\omega) > 0$ for $\omega \in \mathbb{I}_{\alpha}$ according to \eqref{SignD}. For $\omega \in \mathbb{I}_{\alpha}$, by dividing both sides of \eqref{logA} by $\gamma D_\alpha(\omega)$, we get
     \begin{align} \label{PEq}
         & \frac{1}{\gamma}-1 \le -\frac{D_{\beta}(\omega)}{D_{\alpha}(\omega)} = J(\omega;\beta),~\omega \in \mathbb{I}_{\alpha}.
     \end{align}
     Taking the infimum from the right-hand side of \eqref{PEq}, we get \eqref{gam1}.

         It can be verified that
     \begin{subequations} \label{HoRu}
         \begin{align}
             & \lim_{\omega \to 0^{+}} -D_{\beta}(\omega) = 0,~\lim_{\omega \to 0^{+}} D_{\alpha}(\omega) = 0,\\
             & \lim_{\omega \to 0^{+}} -\frac{d D_{\beta}(\omega)}{d \omega} = 0,~\lim_{\omega \to 0^{+}} \frac{d D_{\alpha}(\omega)}{d \omega} = 0,\\
             & \lim_{\omega \to 0^{+}} -\frac{d^2 D_{\beta}(\omega)}{d \omega^2} = \frac{\Delta_{\beta}}{\beta_1^2},~\lim_{\omega \to 0^{+}} \frac{d^2 D_{\alpha}(\omega)}{d \omega^2} = \frac{\Delta_{\alpha}}{\alpha_1^2},
         \end{align}
     \end{subequations}
     hold. According to \eqref{HoRu} and applying L'Hôpital's rule \cite{thomas1992calculus}, we get
     \begin{align} \label{lim1}
         & \lim_{\omega \to 0^{+}} J(\omega;\beta) = \frac{\lim_{\omega \to 0^{+}} -\frac{d^2 D_{\beta}(\omega)}{d \omega^2}}{\lim_{\omega \to 0^{+}} \frac{d^2 D_{\alpha}(\omega)}{d \omega^2}} = \frac{\alpha_1^2}{-\Delta_{\alpha}} \frac{\Delta_{\beta}}{\beta_1^2}.
     \end{align}
     We also have
         \begin{align} \label{lim2}
         & \lim_{\omega \to \sqrt{-\Delta_{\alpha}}^{-}} J(\omega;\beta) = \frac{\lim_{\omega \to \sqrt{-\Delta_{\alpha}}^{-}} -D_{\beta}(\omega)}{\lim_{\omega \to \sqrt{-\Delta_{\alpha}}^{-}} D_{\alpha}(\omega)} = \infty.
     \end{align}
     Considering the limiting behavior of $J(\omega;\beta)$ on interval boundaries (i.e., \eqref{lim1} and \eqref{lim2}) and critical points of $J(\omega;\beta)$, and constructing $\mathcal{J}$ defined by \eqref{CalJ}, $J^{\ast}(\beta)$ in \eqref{gam1} can be computed via \eqref{Jb}.

\section{Proof of Proposition \ref{Propo2}} \label{App4}

     Imposing box constraints \eqref{LUB} to the parameterization \eqref{pqr}, we get
     \begin{subequations} \label{betPar}
        \begin{align}
         &\beta_3^l \le p \le \beta_3^u,\\
         &\beta_2^l \le p+q \le \beta_2^u,\\
         &\beta_1^l \le pq + \frac{q^2}{2} - r \le \beta_1^u
         \end{align}    
     \end{subequations}We have $r \ge 0$. Also, we can consider $p > 0$ and $q > 0$ as $p \ge \epsilon$ and $q \ge \epsilon$, respectively, for an infinitesimal $\epsilon > 0$. Then, \eqref{betPar} along with $p \ge \epsilon$, $q \ge \epsilon$, and $r \ge 0$ implies that
     \begin{subequations} \label{ParProof}
         \begin{align}
            \epsilon \le &p, \label{p1}\\
             \beta_3^l \le &p, \label{p2}\\
             &p \le \beta_3^u, \label{p3}\\
             &p \le \beta_2^u - \epsilon, \label{p4}\\
             &p \le \sqrt{{\beta_2^u}^2-2\beta_1^l}, \label{p5}\\
             \epsilon \le &q, \label{q1}\\
             \beta_2^l-p \le &q, \label{q2}\\
             \sqrt{p^2+2\beta_1^l}-p \le &q, \label{q3}\\
             &q \le \beta_2^u - p, \label{q4}\\
             0 \le &r, \label{r1}\\
             pq + \frac{q^2}{2} - \beta_1^u \le &r, \label{r2}\\
             &r \le pq + \frac{q^2}{2} - \beta_1^l, \label{r3}
         \end{align}
     \end{subequations}hold. Notice that \eqref{q3} is obtained from the combination of \eqref{r1} and \eqref{r3}. Precisely, imposing the non-negativity of the quadratic polynomial $\frac{1}{2}q^2 + pq - \beta_1^l$ subject to $p > 0$, $q > 0$, and $\beta_1^l > 0$, we observe that $\frac{1}{2}q^2 + pq - \beta_1^l \ge 0$ holds if and only if $q \ge \sqrt{p^2+2\beta_1^l}-p$ holds. Also, \eqref{p4} is obtained from the combination of \eqref{q1} and \eqref{q4}. Similarly, \eqref{p5} is obtained from the combination of \eqref{q3} and \eqref{q4}. Thus, utilizing \eqref{ParProof}, the $p$, $q$, and $r$ in \eqref{pqr} satisfying the box constraints \eqref{LUB} can be parameterized as \eqref{ParPar}. Moreover, we have
     \begin{align*}
         & \eqref{p1}, \eqref{p3} \implies \beta_3^u \ge \epsilon,\\
         & \eqref{p1}, \eqref{p4} \implies \beta_2^u \ge 2 \epsilon,\\
         & \eqref{p2}, \eqref{p4} \implies \beta_2^u \ge \beta_3^l + \epsilon,\\
         & \eqref{p2}, \eqref{p5} \implies \beta_2^u \ge \sqrt{{\beta_3^l}^2+2\beta_1^l},
     \end{align*}
     which completes the proof.

\section{An Alternative Approach} \label{App5}

Alternatively, built upon the parameterization \eqref{Partheta} and defining
\begin{align} \label{Ktf}
    & K_{N_{\mathrm{HV}},N_{\mathrm{AV}}}(s;\beta) := F(s;\alpha)^{N_{\mathrm{HV}}}G(s;\beta)^{N_{\mathrm{AV}}},    
\end{align}for any $\beta$ satisfying the stability condition \eqref{DelB}, we can construct the following $\mathcal{H}_{\infty}$-based optimization problem:
\begin{align} \label{objfconst}
     & V^{\ast}(\beta) := \min \{N_{\mathrm{AV}}: \|K_{N_{\mathrm{HV}},N_{\mathrm{AV}}}(s;\beta)\|_{\infty} \le 1\},
\end{align}for a given value of $N_{\mathrm{HV}}$. Then, we can compute the optimal minimum number of the required AVs to stabilize traffic flow by solving the following optimization problem:
\begin{align} \label{gammaf}
     V^{\ast \ast} := \underset{\theta \in \mathbb{R}^3}{\min}~V^{\ast}(\beta(\theta)),
\end{align} for $\tilde{\theta}$. To compute $V^{\ast}(\beta)$ in \eqref{objfconst}, we can utilize a bisection method. To implement \eqref{objfconst}, we can utilize the MATLAB built-in functions $\texttt{tf}()$ and $\texttt{getPeakGain}()$ (developed built upon \cite{bruinsma1990fast}). Note that since $\texttt{getPeakGain}()$ cannot compute the $\mathcal{H}_{\infty}$ norm of $H_{\gamma}(s)$ in \eqref{Hgam}, we are unable to choose $H_{\gamma}(s)$ over $K_{N_{\mathrm{HV}},N_{\mathrm{AV}}}(s;\beta)$ in \eqref{Ktf}. To solve \eqref{gammaf} for $\tilde{\theta}$, we can utilize $\texttt{fminsearch}()$ as our nonlinear optimization solver. Then, we can compute $V^{\ast \ast}$ via $V^{\ast \ast} = V^{\ast}(\beta(\tilde{\theta}))$. Note that $V^{\ast \ast} = \lceil{ {N_{\mathrm{HV}}}/{J^{\ast \ast}}} \rceil$
holds. Equivalently, for a given value of $N_{\mathrm{AV}}$, a similar approach can be used to compute the optimal maximum number of HVs for which traffic flow can be stabilized. Unlike $J^{\ast}(\beta)$ in \eqref{gam1} that takes real values, observe that $V^{\ast}(\beta)$ in \eqref{objfconst} takes positive integer values and as a result, solving \eqref{gammaf} for $\tilde{\theta}$ becomes more challenging and it may rely on the quality of the initialization.

\section{String Stability Conservatism} \label{App6}

We emphasize that all the theoretical derivations in this paper are built upon the string stability criterion \eqref{logfor} equivalently obtained from a sufficient string stability condition $\mathcal{C} \subset \mathbb{C}^{-}$. Consequently, the resulting theoretical bounds have a level of conservatism. In \cite{wu2018stabilizing}, for a single-lane highway setup with a single AV, the authors have chosen a string stability condition $\|\underset{i}{\prod} T_i(s) \|_{\infty} \le 1$ as a formal mathematical definition of string stability (different from the eigenmode-based string stability definition in Definition \ref{def1}). Although such a string stability condition is similar to $\|H_{\gamma}(s)\|_{\infty} \le 1$ associated with the string stability criterion \eqref{logfor}, no conservatism occurs for a single-lane highway setup unlike a circular road setup considered in this paper as $\|\underset{i}{\prod} T_i(s) \|_{\infty} \le 1$ is not a sufficient condition for a single-lane highway setup.

Since the theoretical derivations are agnostic to the distribution of the AVs in the mixed vehicular platoon, one expects to obtain string stability regardless of the distribution of the AVs. However, for the case of $m > 1$, the choice of the distribution of the AVs can potentially affect the quality of the obtained string stability. To quantify such a quality, one can define the following quantity as an $\mathcal{H}_{\infty}$-based measure:
\begin{align*}
    & \chi(\mathcal{I}_{\mathrm{AV}}) := \sum_{j=2}^{n-1} \sum_{i = 1}^{n} \|\prod_{k=i}^{1+\mathrm{mod}(j+i-2,n)} T_k(s) \|_{\infty}.
\end{align*}
Note that for $k \in \mathcal{I}_{\mathrm{AV}}$ and $k \notin \mathcal{I}_{\mathrm{AV}}$, we have $T_k(s) = G(s;\beta)$ and $T_k(s) = F(s;\alpha)$, respectively.

Moreover, considering the following optimization problem:
\begin{align} \label{GrAl}
    \underset{\mathcal{I}_{\mathrm{AV}} \subset \mathbb{N}_n, |\mathcal{I}_{\mathrm{AV}}| = m}{\min}~\chi(\mathcal{I}_{\mathrm{AV}}),
\end{align}
one can search for a sub-optimal distribution of the AVs, namely $\mathcal{I}_{\mathrm{AV}}^{\mathrm{sub-optimal}}$, via employing a greedy algorithm to reduce the string stability conservatism.

\section{Additional Numerical Simulations} \label{App7}

\begin{figure}[H]
    \centering
    \includegraphics[scale = 0.21]{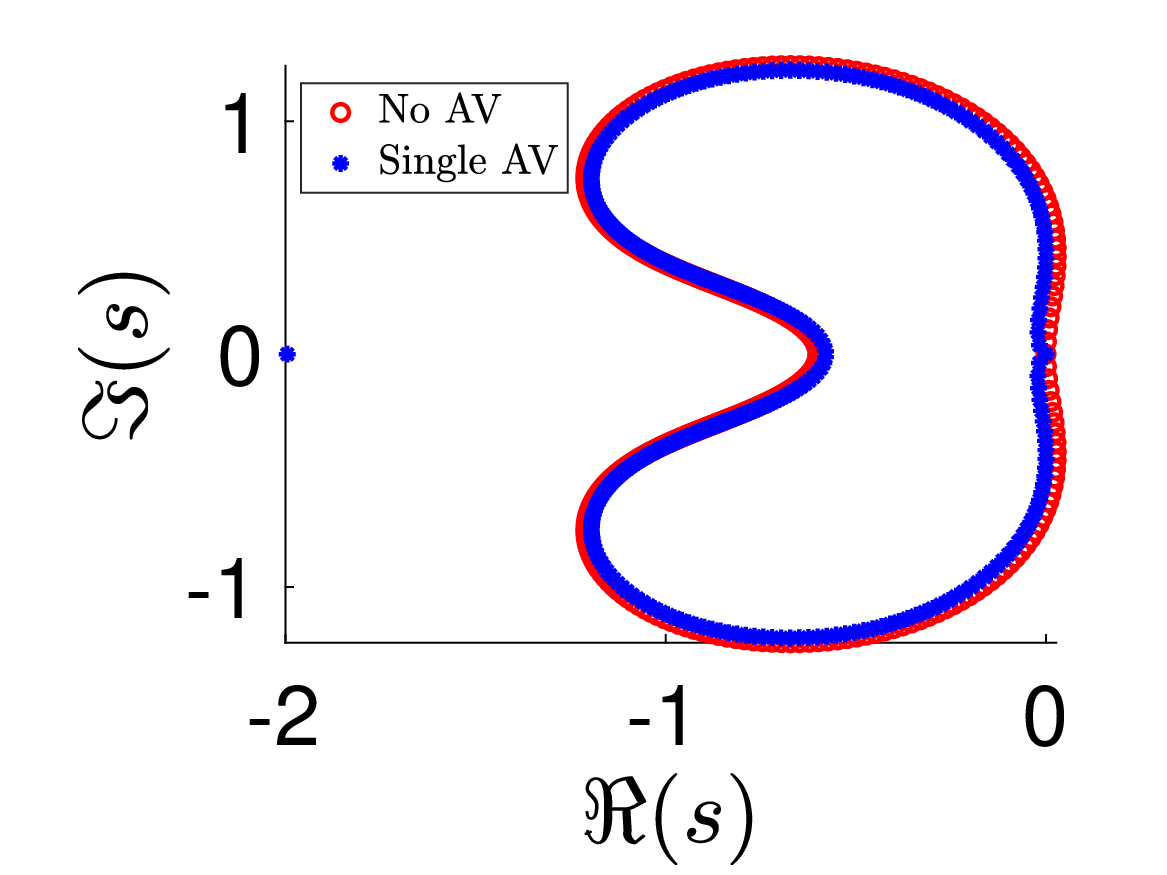}~
    \includegraphics[scale = 0.21]{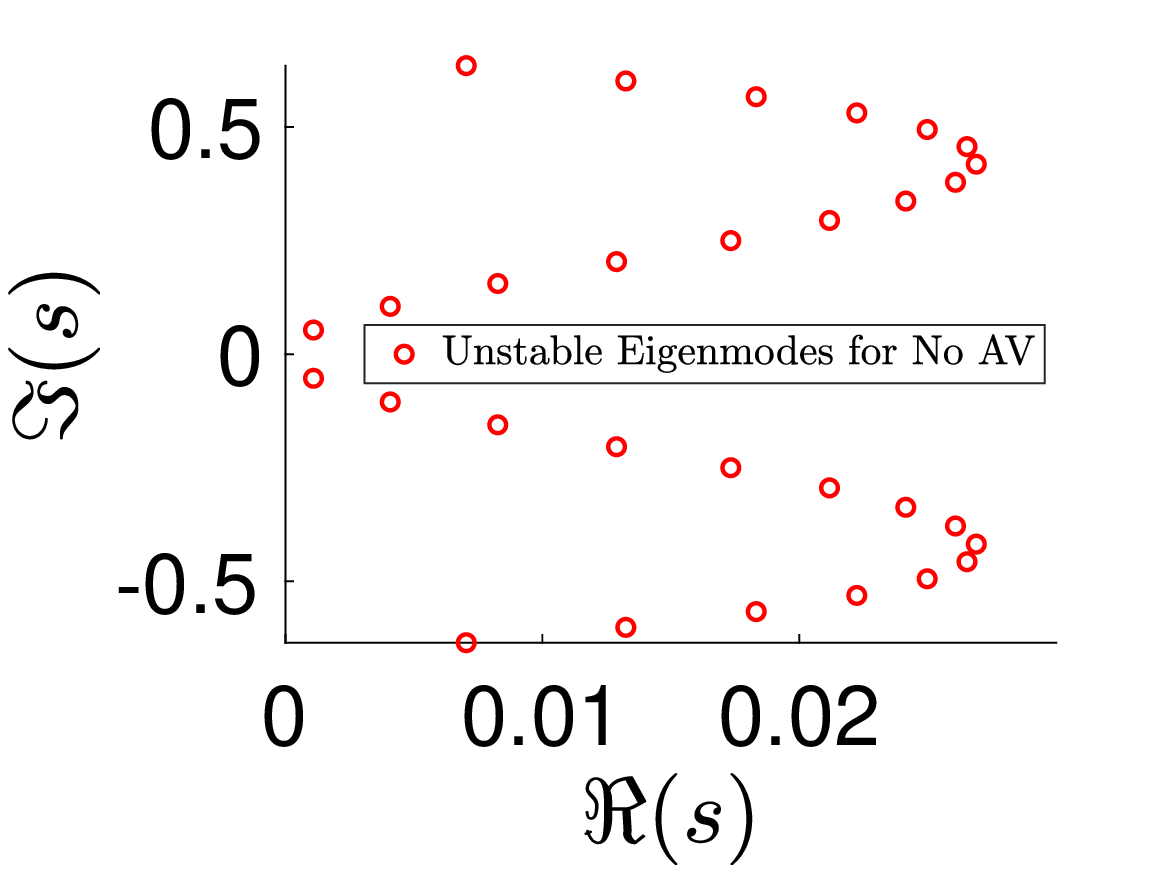}
    \caption{The visualizations of the eigenmodes for the scenarios (i) No AV ($m = 0$), and (ii) Single AV ($m = 1$) with $n = 185$.}
    \label{fig1}
\end{figure}

To empirically validate Corollaries \ref{cor1} and \ref{cor2}, considering the eigenmode-based definition of the string stability in Definition \ref{def1}, we include the visualizations of the eigenmodes for the scenarios (i) No AV ($m = 0$), and (ii) Single AV ($m = 1$) with $n = 185$ in Fig. \ref{fig1}. As Fig. \ref{fig1} demonstrates, the platoon with no AV is string unstable (due to the $15$ paired unstable eigenmodes strictly located on the right half plane) while replacing one of the HVs with a single AV (optimally designed by Procedure $1$) has successfully stabilized the platoon (because all the eigenmodes lie on the left half plane).

\begin{figure}[H]
    \centering
    \includegraphics[scale = 0.21]{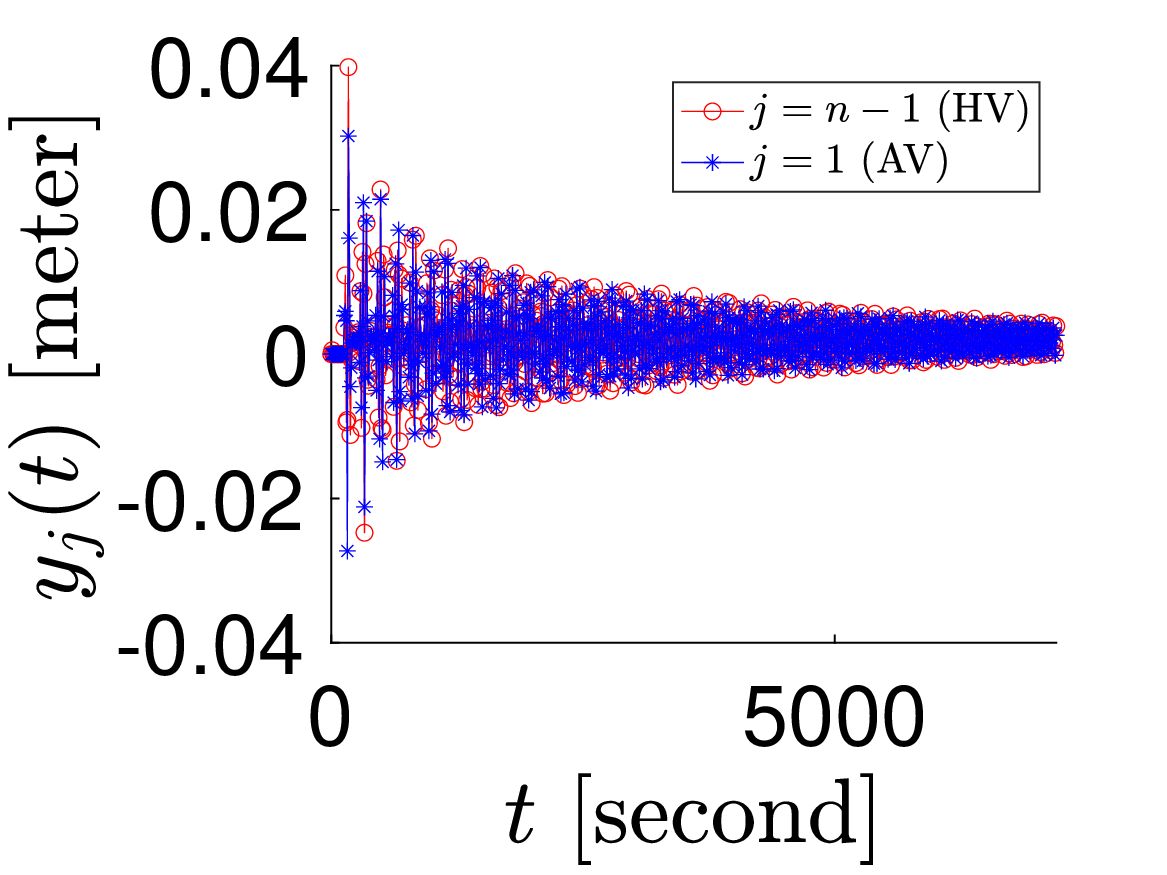}~
    \includegraphics[scale = 0.21]{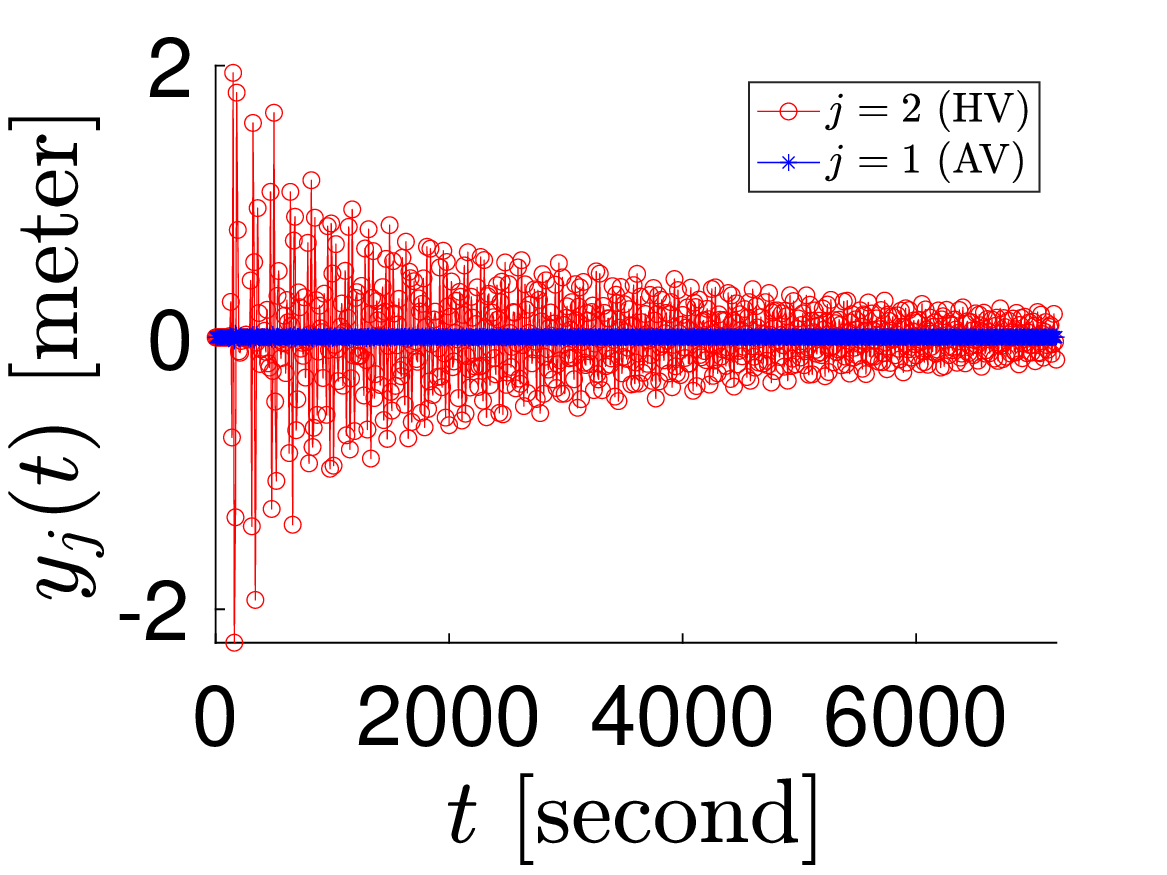}
    \caption{The location deviation trajectories of the $n-1$-th and the $2$-nd vehicles (HVs) and the $1$-st vehicle (AV) for the case of a single AV considered by Fig. \ref{fig1}. (An initial perturbation of magnitude $1$ at $n$-th vehicle's location (HV) is applied.)}
    \label{fig2}
\end{figure}

To showcase the stability obtained by utilizing the single AV, running the simulations with an initial perturbation of magnitude $1$ at $n$-th vehicle's location (HV), we plot the location deviation trajectories of the $(n-1)$-th and the $2$-nd vehicles (HVs) and the $1$-st vehicle (AV) in Fig. \ref{fig2} for the case of a single AV considered by Fig. \ref{fig1}.

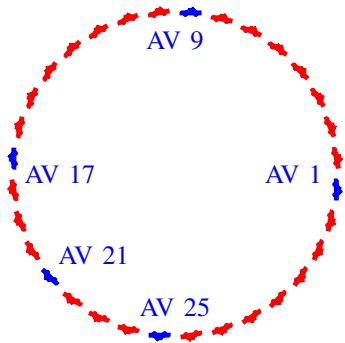
\begin{figure}[H]
    \centering

\begin{tikzpicture} [scale = 0.3]

\tikzset{HV/.pic= {\shade[top color=red, bottom color=red, shading angle={0}]
        [draw=red,fill=red,rounded corners=0ex, thick] (1.45,.5) -- ++(0,0.45) -- ++(0.3,0.3) --  ++(4,0) -- ++(1,0) -- ++(0,-0.75) -- (1.45,.5) -- cycle;
        
    \draw[draw=red, thick, rounded corners=0.5ex,fill=red, thick]  (2.75,1.25) -- ++(0.65,0.65) -- ++(1.6,0) -- ++(0.65,-0.65) -- (2.75,1.25);
    
    \draw[draw=red,,thick]  (4.2,1.25) -- (4.2,1.9);
    
    \draw[draw=red,fill=red,thick] (2.75,.5) circle (.4);
    \draw[draw=red,fill=red,thick] (5.5,.5) circle (.4);
    \draw[draw=red,fill=red,thick] (2.75,.5) circle (.325);
    \draw[draw=red,fill=red,thick] (5.5,.5) circle (.325);}
}

\tikzset{AV/.pic={\shade[top color=blue, bottom color=blue, shading angle={0}]
        [draw=blue,fill=blue,rounded corners=0ex, thick] (1.45,.5) -- ++(0,0.45) -- ++(0.3,0.3) --  ++(4,0) -- ++(1,0) -- ++(0,-0.75) -- (1.45,.5) -- cycle;
        
    \draw[draw=blue, thick, rounded corners=0.5ex,fill=blue, thick]  (2.75,1.25) -- ++(0.65,0.65) -- ++(1.6,0) -- ++(0.65,-0.65) -- (2.75,1.25);
    
    \draw[draw=blue,thick]  (4.2,1.25) -- (4.2,1.9);
    
    \draw[draw=blue,fill=blue,thick] (2.75,.5) circle (.4);
    \draw[draw=blue,fill=blue,thick] (5.5,.5) circle (.4);
    \draw[draw=blue,fill=blue,thick] (2.75,.5) circle (.325);
    \draw[draw=blue,fill=blue,thick] (5.5,.5) circle (.325);}}

 \path (7,0) pic  [rotate=270,scale =0.05] {AV} % 1
 
 ({7*cos(360/32)},{7*sin(360/32)}) pic [rotate={270+360/32},scale =0.05] {HV} % 2
 
 ({7*cos(2*360/32)},{7*sin(2*360/32)}) pic [rotate={270+2*360/32},scale =0.05] {HV} % 3

 ({7*cos(3*360/32)},{7*sin(3*360/32)}) pic [rotate={270+3*360/32},scale =0.05] {HV} % 4

 ({7*cos(4*360/32)},{7*sin(4*360/32)}) pic [rotate={270+4*360/32},scale =0.05] {HV} % 5

 ({7*cos(5*360/32)},{7*sin(5*360/32)}) pic [rotate={270+5*360/32},scale =0.05] {HV} % 6

 ({7*cos(6*360/32)},{7*sin(6*360/32)}) pic [rotate={270+6*360/32},scale =0.05] {HV} % 7

 ({7*cos(7*360/32)},{7*sin(7*360/32)}) pic [rotate={270+7*360/32},scale =0.05] {HV} % 8

 ({7*cos(8*360/32)},{7*sin(8*360/32)}) pic [rotate={270+8*360/32},scale =0.05] {AV} % 9

 ({7*cos(9*360/32)},{7*sin(9*360/32)}) pic [rotate={270+9*360/32},scale =0.05] {HV} % 10
 
 ({7*cos(10*360/32)},{7*sin(10*360/32)}) pic [rotate={270+10*360/32},scale =0.05] {HV} % 11

 ({7*cos(11*360/32)},{7*sin(11*360/32)}) pic [rotate={270+11*360/32},scale =0.05] {HV} % 12

 ({7*cos(12*360/32)},{7*sin(12*360/32)}) pic [rotate={270+12*360/32},scale =0.05] {HV} % 13

 ({7*cos(13*360/32)},{7*sin(13*360/32)}) pic [rotate={270+13*360/32},scale =0.05] {HV} % 14

 ({7*cos(14*360/32)},{7*sin(14*360/32)}) pic [rotate={270+14*360/32},scale =0.05] {HV} % 15

 ({7*cos(15*360/32)},{7*sin(15*360/32)}) pic [rotate={270+15*360/32},scale =0.05] {HV} % 16

 ({7*cos(16*360/32)},{7*sin(16*360/32)}) pic [rotate={270+16*360/32},scale =0.05] {AV} % 17

 ({7*cos(17*360/32)},{7*sin(17*360/32)}) pic [rotate={270+17*360/32},scale =0.05] {HV} % 18
 
 ({7*cos(18*360/32)},{7*sin(18*360/32)}) pic [rotate={270+18*360/32},scale =0.05] {HV} % 19

 ({7*cos(19*360/32)},{7*sin(19*360/32)}) pic [rotate={270+19*360/32},scale =0.05] {HV} % 20

 ({7*cos(20*360/32)},{7*sin(20*360/32)}) pic [rotate={270+20*360/32},scale =0.05] {AV} % 21

 ({7*cos(21*360/32)},{7*sin(21*360/32)}) pic [rotate={270+21*360/32},scale =0.05] {HV} % 22

 ({7*cos(22*360/32)},{7*sin(22*360/32)}) pic [rotate={270+22*360/32},scale =0.05] {HV} % 23

 ({7*cos(23*360/32)},{7*sin(23*360/32)}) pic [rotate={270+23*360/32},scale =0.05] {HV} % 24

 ({7*cos(24*360/32)},{7*sin(24*360/32)}) pic [rotate={270+24*360/32},scale =0.05] {AV} % 25

 ({7*cos(25*360/32)},{7*sin(25*360/32)}) pic [rotate={270+25*360/32},scale =0.05] {HV} % 26
 
 ({7*cos(26*360/32)},{7*sin(26*360/32)}) pic [rotate={270+26*360/32},scale =0.05] {HV} % 27

 ({7*cos(27*360/32)},{7*sin(27*360/32)}) pic [rotate={270+27*360/32},scale =0.05] {HV} % 28

 ({7*cos(28*360/32)},{7*sin(28*360/32)}) pic [rotate={270+28*360/32},scale =0.05] {HV} % 29

 ({7*cos(29*360/32)},{7*sin(29*360/32)}) pic [rotate={270+29*360/32},scale =0.05] {HV} % 30

 ({7*cos(30*360/32)},{7*sin(30*360/32)}) pic [rotate={270+30*360/32},scale =0.05] {HV} % 31

 ({7*cos(31*360/32)},{7*sin(31*360/32)}) pic [rotate={270+31*360/32},scale =0.05] {HV}; % 32

\node[color=blue,scale =1] at (5.25,0) {AV 1};

\node[above,color=blue,scale =1] at ({5.1*cos(8*360/32)},{5.1*sin(8*360/32)}) {AV 9};

\node[color=blue,scale =1] at ({5.1*cos(16*360/32)},{5.1*sin(16*360/32)}) {AV 17};

\node[color=blue,scale =1] at ({5.1*cos(20*360/32)},{5.1*sin(20*360/32)}) {AV 21};

\node[below,color=blue,scale =1] at ({5.1*cos(24*360/32)},{5.1*sin(24*360/32)}) {AV 25};

\end{tikzpicture}
    
    \caption{The visualization of the sub-optimal distribution of the AVs $\mathcal{I}_{\mathrm{AV}}^{\mathrm{sub-optimal}} = \{1,9,17,21,25\}$. $\mathrm{HV}$: red, $\mathrm{AV}$: blue.}
    \label{figg}
\end{figure}

Running Procedure 1 with $\alpha = \begin{bmatrix} 0.3 \pi & 1.5 & 0.9 \end{bmatrix}^\top$, $\beta^l = \begin{bmatrix}
    0.8 & 0.8 & 0.8
\end{bmatrix}$, $\beta^u = \begin{bmatrix}
    2 & 2 & 2
\end{bmatrix}$, we get $\beta(\theta^{\ast}) = \begin{bmatrix}
        0.8 & 2 & 0.8
\end{bmatrix}^\top$ for which $J^{\ast \ast} = 5.4898$ and $\frac{1}{J^{\ast \ast}+1} = 0.1541$. Choosing $N_{\mathrm{AV}} = 5$, (23b) implies that $N_{\mathrm{HV}} \le \lfloor 5.4898 \times 5 \rfloor = 27$. Then, choosing $(m,n) = (5,32)$ and solving \eqref{GrAl} via a greedy algorithm, we obtain the sub-optimal distribution of the AVs $\mathcal{I}_{\mathrm{AV}}^{\mathrm{sub-optimal}} = \{1,9,17,21,25\}$ visualized by Fig. \ref{figg}.

\end{document}

%% file: preamble.tex
\usepackage{lcsys}
\usepackage{epsfig,color,amsmath,cite}
\IEEEoverridecommandlockouts
\usepackage{empheq}
\usepackage{bm}
\usepackage{epstopdf}
\usepackage{amssymb}
\usepackage{url}
\usepackage{multirow}
\usepackage{hhline}
\usepackage{booktabs}

\usepackage{graphicx}
\usepackage{algorithmic}

\usepackage[linesnumbered,boxed,commentsnumbered,ruled,vlined,longend]{algorithm2e}

\SetAlgorithmName{Procedure}

\newtheorem{mydef}{Definition}

\newtheorem{mypbm}{Problem}

\newtheorem{mycor}{Corollary}
\newtheorem{myprs}{Proposition}



\usepackage{stackengine}

\allowdisplaybreaks[4]
\usepackage[colorlinks = true,
linkcolor = blue,
urlcolor  = blue,
citecolor = blue,
anchorcolor = blue]{hyperref}

\usepackage[framemethod=TikZ]{mdframed}
\mdfdefinestyle{MyFrame}{%
	linecolor=black,
	outerlinewidth=1.25pt,
	roundcorner=1.25pt,
	innerrightmargin=5pt,
	innerleftmargin=5pt,}
	
\usepackage[noabbrev]{cleveref}

\usepackage{mathtools}

\DeclarePairedDelimiter\abs{\lvert}{\rvert}%
\DeclarePairedDelimiter\norm{\lVert}{\rVert}%

\makeatletter
\let\oldabs\abs
\def\abs{\@ifstar{\oldabs}{\oldabs*}}
\let\oldnorm\norm
\def\norm{\@ifstar{\oldnorm}{\oldnorm*}}
\makeatother

\usepackage[english]{babel}
\usepackage[utf8]{inputenc}
\usepackage[super]{nth}

\usepackage{graphicx}
\usepackage{float}

\usepackage{array}
\usepackage{threeparttable}

\usepackage[english]{babel}
\usepackage[utf8]{inputenc}
\usepackage[super]{nth}

\usepackage{pifont}% http://ctan.org/pkg/pifont

%% file: root_arXiv.bbl
\begin{thebibliography}{10}
\providecommand{\url}[1]{#1}
\csname url@samestyle\endcsname
\providecommand{\newblock}{\relax}
\providecommand{\bibinfo}[2]{#2}
\providecommand{\BIBentrySTDinterwordspacing}{\spaceskip=0pt\relax}
\providecommand{\BIBentryALTinterwordstretchfactor}{4}
\providecommand{\BIBentryALTinterwordspacing}{\spaceskip=\fontdimen2\font plus
\BIBentryALTinterwordstretchfactor\fontdimen3\font minus
  \fontdimen4\font\relax}
\providecommand{\BIBforeignlanguage}[2]{{%
\expandafter\ifx\csname l@#1\endcsname\relax
\typeout{** WARNING: IEEEtran.bst: No hyphenation pattern has been}%
\typeout{** loaded for the language `#1'. Using the pattern for}%
\typeout{** the default language instead.}%
\else
\language=\csname l@#1\endcsname
\fi
#2}}
\providecommand{\BIBdecl}{\relax}
\BIBdecl

\bibitem{cui2017stabilizing}
S.~Cui, B.~Seibold, R.~Stern, and D.~B. Work, ``Stabilizing traffic flow via a
  single autonomous vehicle: Possibilities and limitations,'' in \emph{IEEE
  Intelligent Vehicles Symposium (IV)}, 2017, pp. 1336--1341.

\bibitem{stern2018dissipation}
R.~E. Stern, S.~Cui, M.~L. Delle~Monache, R.~Bhadani, M.~Bunting, M.~Churchill,
  N.~Hamilton, H.~Pohlmann, F.~Wu, B.~Piccoli \emph{et~al.}, ``Dissipation of
  stop-and-go waves via control of autonomous vehicles: Field experiments,''
  \emph{Transportation Research Part C: Emerging Technologies}, vol.~89, pp.
  205--221, 2018.

\bibitem{wu2018stabilizing}
C.~Wu, A.~M. Bayen, and A.~Mehta, ``Stabilizing traffic with autonomous
  vehicles,'' in \emph{IEEE International Conference on Robotics and Automation
  (ICRA)}, 2018, pp. 6012--6018.

\bibitem{zheng2020smoothing}
Y.~Zheng, J.~Wang, and K.~Li, ``Smoothing traffic flow via control of
  autonomous vehicles,'' \emph{IEEE Internet of Things Journal}, vol.~7, no.~5,
  pp. 3882--3896, 2020.

\bibitem{wang2021controllability}
J.~Wang, Y.~Zheng, Q.~Xu, J.~Wang, and K.~Li, ``Controllability analysis and
  optimal control of mixed traffic flow with human-driven and autonomous
  vehicles,'' \emph{IEEE Transactions on Intelligent Transportation Systems},
  vol.~22, no.~12, pp. 7445--7459, 2021.

\bibitem{wang2021optimal}
S.~Wang, R.~Stern, and M.~W. Levin, ``Optimal control of autonomous vehicles
  for traffic smoothing,'' \emph{IEEE Transactions on Intelligent
  Transportation Systems}, vol.~23, no.~4, pp. 3842--3852, 2021.

\bibitem{wang2023optimal}
S.~Wang, M.~W. Levin, and R.~Stern, ``Optimal feedback control law for
  automated vehicles in the presence of cyberattacks: A min--max approach,''
  \emph{Transportation research part C: emerging technologies}, vol. 153, p.
  104204, 2023.

\bibitem{wang2023general}
S.~Wang, M.~Shang, M.~W. Levin, and R.~Stern, ``A general approach to smoothing
  nonlinear mixed traffic via control of autonomous vehicles,''
  \emph{Transportation Research Part C: Emerging Technologies}, vol. 146, p.
  103967, 2023.

\bibitem{ameli2024designing}
M.~Ameli, S.~McQuade, J.~W. Lee, M.~Bunting, M.~Nice, H.~Wang, W.~Barbour,
  R.~Weightman, C.~Denaro, R.~Delorenzo \emph{et~al.}, ``Designing, simulating,
  and performing the 100-{AV} field test for the circles consortium:
  Methodology and implementation of the largest mobile traffic control
  experiment to date,'' \emph{arXiv preprint arXiv:2404.15533}, 2024.

\bibitem{zhou2020stabilizing}
Y.~Zhou, S.~Ahn, M.~Wang, and S.~Hoogendoorn, ``Stabilizing mixed vehicular
  platoons with connected automated vehicles: An {H}-infinity approach,''
  \emph{Transportation Research Part B: Methodological}, vol. 132, pp.
  152--170, 2020.

\bibitem{bahavarnia2024on}
M.~Bahavarnia, J.~Ji, A.~F. Taha, and D.~B. Work, ``On the {CAV} platoon
  control problem: Constrained vs unconstrained,'' in \emph{63rd IEEE
  Conference on Decision and Control (CDC)}.\hskip 1em plus 0.5em minus
  0.4em\relax IEEE, 2024, pp. 1--8.

\bibitem{wilson2011car}
R.~E. Wilson and J.~A. Ward, ``Car-following models: fifty years of linear
  stability analysis--a mathematical perspective,'' \emph{Transportation
  Planning and Technology}, vol.~34, no.~1, pp. 3--18, 2011.

\bibitem{swaroop1996string}
D.~Swaroop and J.~K. Hedrick, ``String stability of interconnected systems,''
  \emph{IEEE Transactions on Automatic Control}, vol.~41, no.~3, pp. 349--357,
  1996.

\bibitem{bando1995dynamical}
M.~Bando, K.~Hasebe, A.~Nakayama, A.~Shibata, and Y.~Sugiyama, ``Dynamical
  model of traffic congestion and numerical simulation,'' \emph{Physical Review
  E}, vol.~51, no.~2, p. 1035, 1995.

\bibitem{chou2022lord}
F.-C. Chou, A.~R. Bagabaldo, and A.~M. Bayen, ``The lord of the ring road: a
  review and evaluation of autonomous control policies for traffic in a ring
  road,'' \emph{ACM Transactions on Cyber-Physical Systems (TCPS)}, vol.~6,
  no.~1, pp. 1--25, 2022.

\bibitem{di2021survey}
X.~Di and R.~Shi, ``A survey on autonomous vehicle control in the era of
  mixed-autonomy: From physics-based to ai-guided driving policy learning,''
  \emph{Transportation research part C: emerging technologies}, vol. 125, p.
  103008, 2021.

\bibitem{wu2021flow}
C.~Wu, A.~R. Kreidieh, K.~Parvate, E.~Vinitsky, and A.~M. Bayen, ``Flow: A
  modular learning framework for mixed autonomy traffic,'' \emph{IEEE
  Transactions on Robotics}, vol.~38, no.~2, pp. 1270--1286, 2021.

\bibitem{hasan2024lessons}
A.~Hasan, N.~Chakraborty, H.~Chen, C.~Wu, and K.~Driggs-Campbell, ``Lessons in
  cooperation: A qualitative analysis of driver sentiments towards real-time
  advisory systems from a driving simulator user study,'' \emph{arXiv preprint
  arXiv:2407.13775}, 2024.

\bibitem{sugiyama2008traffic}
Y.~Sugiyama, M.~Fukui, M.~Kikuchi, K.~Hasebe, A.~Nakayama, K.~Nishinari, S.-i.
  Tadaki, and S.~Yukawa, ``Traffic jams without bottlenecks—experimental
  evidence for the physical mechanism of the formation of a jam,'' \emph{New
  journal of physics}, vol.~10, no.~3, p. 033001, 2008.

\bibitem{bando1994structure}
M.~Bando, K.~Hasebe, A.~Nakayama, A.~Shibata, and Y.~Sugiyama, ``Structure
  stability of congestion in traffic dynamics,'' \emph{Japan Journal of
  Industrial and Applied Mathematics}, vol.~11, pp. 203--223, 1994.

\bibitem{lagarias1998convergence}
J.~C. Lagarias, J.~A. Reeds, M.~H. Wright, and P.~E. Wright, ``Convergence
  properties of the {Nelder}--{Mead} simplex method in low dimensions,''
  \emph{SIAM Journal on Optimization}, vol.~9, no.~1, pp. 112--147, 1998.

\bibitem{li2024hybrid}
S.~Li, R.~Dong, and C.~Wu, ``Hybrid system stability analysis of multi-lane
  mixed-autonomy traffic,'' \emph{IEEE Transactions on Robotics}, 2024.

\bibitem{thomas1992calculus}
G.~B. Thomas, ``Calculus and analytic geometry,'' \emph{Massachusetts Institute
  of Technology, Massachusetts, USA, Addison-Wesley Publishing Company, ISBN:
  0-201-60700-X}, 1992.

\bibitem{bruinsma1990fast}
N.~Bruinsma and M.~Steinbuch, ``A fast algorithm to compute the
  $\mathcal{H}_{\infty}$-norm of a transfer function matrix,'' \emph{Systems \&
  Control Letters}, vol.~14, no.~4, pp. 287--293, 1990.

\end{thebibliography}
